\documentstyle[aps,preprint,epsfig]{revtex}
\begin{document}
%\begin{fmffile}{fmfphysrevbosefermi}

\draft
\title{
Quantum field theory of dilute homogeneous Bose-Fermi \\
mixtures at zero temperature: general formalism \\ 
and beyond mean-field corrections}
\author{A. P. Albus,$^{1}$ S. A. Gardiner,$^{1,2}$ F. Illuminati,$^{3,4}$ 
and M. Wilkens$^{1}$}
\address{\mbox{}$^{1}$Institut f\"{u}r Physik, Universit\"{a}t Potsdam, 
D--14469 Potsdam, Germany\\
\mbox{}$^{2}$Institut f\"{u}r Theoretische Physik, Universit\"{a}t 
Hannover, D--30167 Hannover, Germany\\
\mbox{}$^{3}$Dipartimento di Fisica, Universit\`{a} degli Studi di Salerno, 
I--84081 Baronissi (SA), Italy\\
\mbox{}$^{4}$Istituto Nazionale di Fisica della Materia (INFM), 
Unit\`{a} di Salerno, Baronissi (SA), Italy}
\date{\today}
\maketitle

\begin{abstract}
We consider a dilute homogeneous mixture of bosons and spin-polarized 
fermions at zero temperature. We first construct the formal
scheme for carrying out systematic perturbation
theory in terms of single particle Green's functions. We
especially focus on the description of the boson-fermion interaction. 
To do so we need to introduce a new relevant object,
the renormalized boson-fermion $T$-matrix which we determine 
to second order in the boson-fermion $s$-wave scattering length. 
We also discuss how to incorporate the usual
boson-boson $T$-matrix in mean-field approximation to obtain the total
ground state properties of the system.
The next order term beyond mean-field stems from the boson-fermion interaction
and is proportional to $a_{\mbox{\scriptsize BF}}k_{\mbox{\scriptsize F}}$.
The total ground-state energy-density reads
$E/V = \epsilon_{\mbox{\scriptsize F}} + \epsilon_{\mbox{\scriptsize B}} +
(2\pi\hbar^{2}a_{\rm  BF}n_{\mbox{\scriptsize B}}n_{\mbox{\scriptsize F}}/m)
[1 + a_{\mbox{\scriptsize BF}}k_{\mbox{\scriptsize F}}f(\delta)/\pi]$.
The first term is the kinetic energy of the free fermions,
the second term is the boson-boson mean-field interaction, 
the pre-factor to the additional term is
the usual mean-field contribution to the boson-fermion
interaction energy, and the second term in the square
brackets is the second-order correction, where $f(\delta)$
is a known function of 
$\delta= (m_{\mbox{\scriptsize B}} 
- m_{\mbox{\scriptsize F}})/(m_{\mbox{\scriptsize B}} 
+ m_{\mbox{\scriptsize F}})$.
We also compute the bosonic and the fermionic chemical
potentials, the compressibilities, and the modification to
the induced fermion-fermion interaction.
We discuss the behavior of the total ground-state energy 
and the importance of the beyond mean-field correction 
for various parameter regimes, in particular considering
mixtures of $^{6}$Li and $^{7}$Li and of $^{3}$He and
$^{4}$He. Moreover we determine the modification of the
induced fermion-fermion interaction due to the beyond
mean-field effects. We show that there is no effect on
the depletion of the Bose condensate to first order in
the boson-fermion scattering length $a_{\mbox{\scriptsize BF}}$.
\end{abstract}


\narrowtext

\section{Introduction}
Following the spectacular success in achieving
Bose-Einstein condensation in trapped, dilute atomic gases in
1995 \cite{AndEns95,DavMew95,BraSac95}, there has been an 
explosion of experimental and theoretical activity on 
this newly accessible state of matter (for recent
reviews focusing on different experimental and
theoretical aspects see for instance
\cite{Dalf99,IngStr99,Ket99,Legg01,Baym01}). 
More recently, there has been
increasing interest and experimental activity also 
in quantum degenerate ultra-cold Fermi gases
\cite{DeMJin99,DeMPap01,GolPap01,TruStr01,SchFer01,SchKha01}, 
in particular because of the possibility of observing a BCS 
type transition in a dilute atomic
gas \cite{StoHou96,HolKok01}. Dilute mixtures of ultra-cold gases
of bosonic 
and fermionic atoms are also receiving increased attention, 
in particular because sympathetic
cooling of the fermions by the 
bosons is an important means of their achieving
quantum degeneracy \cite{GolPap01,TruStr01,SchFer01,SchKha01},
and also because bosons can mediate an induced (attractive)
fermion-fermion interaction \cite{VivPet00}.
Moreover mixtures of atomic $^{3}$He and $^{4}$He have become
interesting in their own right after the recent achievement of 
Bose-Einstein condensation in metastable $^{4}$He
\cite{RobSir01,PerLeo01}, as they could represent a bridge towards
the understanding of superfluidity in helium.

Current analyses of dilute mixtures of ultra-cold atomic 
boson and fermion vapors
are based on mean-field approximations. They include
for example the work on  
stability considerations for homogeneous systems by 
Viverit {\em et al}.\
\cite{VivPet00}, and the calculation of density 
distributions and phase separation of trapped mixtures
by Nygaard and M{\o}lmer \cite{NygMol98}.
Some interesting
effects have been studied
by Bijlsma {\em et al}.\ \cite{BijHer00},
where an effective modification of the 
fermion-fermion scattering length, 
mediated by boson-fermion scattering 
processes, was determined.
Pu {\em et al}.\ determined the phonon spectrum of
the Bose condensate in a boson-fermion mixture at
zero temperature \cite{PuZha01}.

Mean-field approaches have proved to be 
extremely useful in the theoretical and
experimental study of
Bose-Einstein condensed dilute 
atomic gases, and are likely to prove similarly
useful for quantum degenerate mixed boson-fermion systems. It is nevertheless
desirable to consider beyond mean-field effects, and under what circumstances
they are likely to be most relevant.
For pure (unpolarized) fermion \cite{Gal58,AbrGor63,FetWal71} and pure boson 
\cite{AbrGor63,FetWal71,LeeYan57,Bel58,HugPin59,Wu59,Saw59,BraNie} systems, 
expansions of the ground-state energy, in terms of the small parameters
$k_{\mbox{\scriptsize F}}a_{\mbox{\scriptsize  FF}}$ and 
$\sqrt{n_{\mbox{\scriptsize B}}a_{\mbox{\scriptsize BB}}^{3}}$ ($k_{\mbox{\scriptsize F}}$ 
is the Fermi wavenumber,
$n_{\mbox{\scriptsize B}}$ the boson density, $a_{\mbox{\scriptsize FF}}$ 
and $a_{\mbox{\scriptsize BB}}$ the 
fermion-fermion and boson-boson scattering
lengths), are well established. These expansions go beyond mean-field 
approximations while still depending only on the $s$-wave 
scattering lengths. Although
determined for homogeneous systems, the use of
beyond mean-field corrections arising from consideration of such
expansions may be readily extended to the experimentally
relevant case of 
inhomogeneous trapped gases by
application of the local density approximation.
In general, the beyond mean-field corrections for the bosons are smaller than
for the fermions, since the exponent of the small dimensionless parameter $n^{1/3}a$
($n$ is the density parameter and $a$ the scattering length) is $1$ in the fermion case
but $3/2$ in the boson case.

In the case of dilute fermions immersed in a Bose
gas an expansion of the ground-state energy in terms of
the small parameters $\sqrt{a_{\mbox{\scriptsize BB}}^{3} n_{\mbox{\scriptsize B}}}$ 
and $n_{\mbox{\scriptsize F}}/n_{\mbox{\scriptsize B}}$,
where $n_{\mbox{\scriptsize F}}$ is the fermion density,
was performed by Saam \cite{Saa69}. 
This was motivated by considering quantum-degenerate dilute gases as a
model for the behavior of 
superfluid helium, where the assumption of $n_{\mbox{\scriptsize F}}/n_{\mbox{\scriptsize B}}$ as a 
small parameter is justified by the much greater 
natural occurrence of bosonic ${}^4$He compared to that of the 
fermionic ${}^3$He isotope.

Systems where there are vastly more bosons than fermions are certainly
experimentally achievable in dilute atomic gases, and it can in fact be
advantageous to have an excess of bosons in order to enhance
sympathetic cooling \cite{GolPap01}. However, there is in principle
no {\em a priori\/} reason to confine theoretical analyses
to such systems. In fact, in recent experiments 
\cite{SchFer01,SchKha01} the numbers of fermions and bosons are 
comparable. Thus motivated, in the present paper we
derive a systematic perturbative expansion for the 
ground-state energy and other related relevant
physical quantities for dilute Bose-Fermi
mixtures at zero temperature and for arbitrary ratios
of the boson and fermion densities.
In this way we determine
the lowest-order correction to mean-field in the case of 
weakly interacting bosons and spin-polarized fermions in terms of the
the gas parameter $k_{\mbox{\scriptsize F}}a_{\mbox{\scriptsize BF}}$, 
where $a_{\mbox{\scriptsize BF}}$ is the boson-fermion
$s$-wave scattering length. 
%This approach is then 
%capable of dealing with situations where the 
%relative concentration of bosons and 
%fermions is arbitrary, provided that
%the two gas parameters remain small.
The ground-state
energy thus derived can then be implemented, in local density 
approximation, as the energy functional for the
study of the experimentally relevant case of
trapped mixtures, in complete analogy
with the pure bosonic and pure fermionic cases.

The plan of the paper is as follows. 
In Section \ref{system} we introduce the basic Hamiltonian
for a system of interacting bosons and spin-polarized
fermions, expressed in its grand-canonical form after
performing the Bogoliubov replacement.
In Section \ref{systematic} we define the one particle
Green's functions needed for a systematic 
field-theoretical analysis of the boson-boson and 
boson-fermion interactions, and we determine the
associated Feynman rules. In Section \ref{determination}
we implement the perturbative expansion by introducing the
boson-fermion self-energy and the
renormalized boson-fermion $T$-matrix in ladder approximation 
and by solving the corresponding Bethe-Salpeter equation
to second order in $k_{\mbox{\scriptsize F}}a_{\mbox{\scriptsize BF}}$.
In Section \ref{results} we exploit the results obtained
in the previous Sections to compute some relevant physical
quantities. In particular we provide the expression for
the ground-state energy density to second order in the
gas parameter, the bosonic and fermionic chemical potentials,
the compressibilities, and the induced fermion-fermion interaction.
We then compare the results thus obtained with actual and
foreseeable experimental situations to assess the relative
importance of higher-order corrections with respect to
the mean-field results. 
In Section \ref{discussion} conclusions are drawn and some
possible future developments are discussed.
 
\section{System}
\label{system}
\subsection{Hamiltonian and ground-state energy}
\label{hamilton}
\subsubsection{Many-body Hamiltonian}
We consider a homogeneous mixture of interacting bosons and fermions,
imposing periodic boundary conditions on a volume $V$. 
In complete generality there are thus
boson-boson, boson-fermion, and fermion-fermion interactions to consider.
However, for spin-polarized fermions, there is no $s$-wave scattering
contribution to the fermion-fermion interaction \cite{GalPas}. The first
non-vanishing contribution is due to $p$-wave scattering, 
which can generally be neglected 
when compared to the boson-boson and boson-fermion interactions, 
which are due to $s$-wave scattering. 
We thus take into account $s$-wave scattering between
bosons, and between bosons and fermions only. 

In second-quantized form, the
Hamiltonian describing this situation is
\begin{equation}
\hat{H}=\hat{T}_{\mbox{\scriptsize B}}
+ \hat{T}_{\mbox{\scriptsize F}}+\hat{U}+\hat{V},
\end{equation}
where
\begin{eqnarray}
\hat{T}_{\mbox{\scriptsize B}}&=&
\frac{\hbar^2}{2m_{\mbox{\scriptsize B}}}
\int d^3{\mathbf{x}}{\mathbf{\nabla}}\hat{\Phi}^\dagger({\mathbf{x}})
\cdot{\mathbf{\nabla}}\hat{\Phi}({\mathbf{x}}), 
\\
\hat{T}_{\mbox{\scriptsize F}}&=&
\frac{\hbar^2}{2m_{\mbox{\scriptsize F}}}
\int d^3{\mathbf{x}}{\mathbf{\nabla}}\hat{\Psi}^\dagger({\mathbf{x}})
\cdot{\mathbf{\nabla}}\hat{\Psi}({\mathbf{x}}), 
\\
\hat{U}&=&
\int \int d^3{\mathbf{x}} d^3{\mathbf{x}}' 
\hat{\Phi}^\dagger({\mathbf{x}}) \hat{\Psi}^\dagger({\mathbf{x}}')
      U(|{\mathbf{x}}-{\mathbf{x}}'|) 
\hat{\Psi}({\mathbf{x}}')\hat{\Phi}({\mathbf{x}}), 
\\
\hat{V}&=&
\frac{1}{2}\int \int d^3{\mathbf{x}} d^3{\mathbf{x}}' 
\hat{\Phi}^\dagger({\mathbf{x}}) \hat{\Phi}^\dagger({\mathbf{x}}')
          V(|{\mathbf{x}}-{\mathbf{x}}'|)
\hat{\Phi}({\mathbf{x}}')\hat{\Phi}({\mathbf{x}}), 
\end{eqnarray}
and where $\hat{\Phi}({\mathbf{x}})$ is a bosonic 
field operator, $\hat{\Psi}({\mathbf{x}})$ is a
fermionic field operator, and 
$m_{\mbox{\scriptsize B}}$ and $m_{\mbox{\scriptsize F}}$ 
are the respective masses of
the bosons and fermions. 
For later reference we also define
\begin{eqnarray}
\hat{H}_0&=&\hat{T}_{\mbox{\scriptsize B}}
+ \hat{T}_{\mbox{\scriptsize F}},\\
\hat{W}&=&\hat{U}+\hat{V}.
\end{eqnarray}

\subsubsection{Mean field theory}
It is straightforward to determine a zero-temperature 
mean field theory for 
$\hat{H}$ \cite{NygMol98}. 
Employing the well-known Thomas-Fermi approximation,
the mean field ground-state energy density is
\begin{equation}
\frac{E}{V}=\frac{3}{5}
\frac{\hbar^2 k_{\mbox{\scriptsize F}}^2}{2m_{\mbox{\scriptsize F}}}n_{\mbox{\scriptsize F}}
+ \frac{2\pi \hbar^2
a_{\mbox{\scriptsize BF}}}{m}n_{\mbox{\scriptsize B}}n_{\mbox{\scriptsize F}}+\frac{2\pi \hbar^2 a_{\mbox{\scriptsize BB}}}{m_{\mbox{\scriptsize B}}}n^2_{\mbox{\scriptsize B}},
\label{mfe}
\end{equation}
where $m=m_{\mbox{\scriptsize F}}m_{\mbox{\scriptsize B}}/(m_{\mbox{\scriptsize F}}+m_{\mbox{\scriptsize B}})$ is the reduced mass, 
and $k_{\mbox{\scriptsize F}}=(6\pi^{2}n_{\mbox{\scriptsize F}})^{1/3}$ is the fermi wave-number \cite{kf}. 
In the case of a pure fermionic system, beyond mean-field
corrections to the ground state energy-density are given by \cite{Gal58}:
\begin{eqnarray}
\frac{E_{\mbox{\scriptsize F}}}{V}=
\frac{3\hbar^2 k_F^2}{5 m_{\mbox{\scriptsize F}}}n_{\mbox{\scriptsize F}}\left[1+
\frac{128}{15}k_{\mbox{\scriptsize F}}a_{\mbox{\scriptsize F}}+
(k_{\mbox{\scriptsize F}}a_{\mbox{\scriptsize F}})^2+\ldots
\right].
\label{EnDen}
\end{eqnarray}
For pure bosons corrections to the ground state have been calculated 
by, e.g., Hugenholtz and Pines
\cite{HugPin59} and by Wu \cite{Wu59}.
These corrections are obtained via a perturbative expansion
in terms of the bosonic gas parameter 
$\sqrt{n_{\mbox{\scriptsize B}}a^3_{\mbox{\scriptsize BB}}}$. 
As already mentioned, this parameter is
in general smaller than the fermionic gas parameter 
(see also Sec.\ \ref{groundE}).
%Beyond the order given in Eq.~(\ref{EnDen}), however,
%corrections to the energy functional generally depend 
%on the form of the interaction potential
%\cite{Bel58,BraNie}, not just the scattering length. 
Our goal is thus to determine a {\em general \/} 
expression equivalent to Eq.~(\ref{EnDen}), taking into
account boson-fermion interactions, while neglecting 
corrections proportional to higher powers of the bosonic
gas parameter.
 
\subsection{Bogoliubov replacement and grand-canonical Hamiltonian}
\label{theory}
\label{bogoliubov}

In order to determine the energy functional to higher order than 
in Eq.\ (\ref{mfe}), we will adopt a
perturbative approach using one-particle Green's functions,
in a way essentially equivalent to the field-theoretical
treatment of pure bosonic and
fermionic systems  \cite{AbrGor63,FetWal71}. 
We thus first carry out the Bogoliubov replacement \cite{Bog47},
where the condensate bosons are treated as a $c$-number field:
\begin{equation}
\hat{\Phi}({\mathbf{x}}) =\sqrt{n_0}+\hat{\phi}({\mathbf{x}}),
 \label{Bogo}
\end{equation}
where $n_0=N_{0}/V$ is the condensate density, and $N_{0}$ is the number of
(condensate)
atoms in the $k=0$ mode. This prescription
breaks particle number conservation (see 
\cite{Arn59,Gar97,CasDum98,Mor00}
for alternative Bogoliubov replacements that
preserve particle number conservation);
average particle number conservation is assured by introducing the
grand-canonical Hamiltonian
\begin{equation}
\hat{K}=\hat{H}-\mu_{\mbox{\scriptsize B}}\hat{N}_{\mbox{\scriptsize B}},
\label{Kam}
\end{equation} 
where $\mu_{\mbox{\scriptsize B}}$ is a Lagrange multiplier, to be identified with the boson chemical potential
\cite{FetWal71}. 
Substituting Eq.~(\ref{Bogo}) into Eq.~(\ref{Kam}),
the grand-canonical Hamiltonian reads:
\begin{equation}
\hat{K}=\hat{K}_0
-\mu_{\mbox{\scriptsize B}} N_{0}+\hat{U}_1+\hat{U}_2+\hat{U}_3
+\hat{V}_1+\hat{V}_2+\hat{V}_3+\hat{V}_4+\hat{V}_5+\hat{V}_6+\hat{V}_7,
\end{equation}
where
\begin{equation}
 \hat{K_0}=\frac{\hbar^2}{2m_{\mbox{\scriptsize B}}}
\int 
d^3{\mathbf{x}}{\mathbf{\nabla}}\hat{\phi}^\dagger({\mathbf{x}})
\cdot{\mathbf{\nabla}}
\hat{\phi}({\mathbf{x}})
+\frac{\hbar^2}{2m_{\mbox{\scriptsize F}}}\int d^3{\mathbf{x}}{\mathbf{\nabla}}\hat{\Psi}^\dagger({\mathbf{x}})
\cdot{\mathbf{\nabla}}\hat{\Psi}({\mathbf{x}})
-\mu_{\mbox{\scriptsize B}}\int d^3{\mathbf{x}}\hat{\phi}^\dagger({\mathbf{x}})\hat{\phi}({\mathbf{x}}),
\label{K0}
\end{equation}
\begin{eqnarray}
\hat{U}_1&=&n_0\int \int d^3{\mathbf{x}} d^3{\mathbf{x}}' \hat{\Psi}^\dagger({\mathbf{x}}')
         U(|{\mathbf{x}}-{\mathbf{x}}'|)\hat{\Psi}({\mathbf{x}}'),\\
\hat{U}_2&=&\sqrt{n_0}\int \int d^3{\mathbf{x}} d^3{\mathbf{x}}' \hat{\Psi}^\dagger({\mathbf{x}}')
            U(|{\mathbf{x}}-{\mathbf{x}}'|)
\hat{\Psi}({\mathbf{x}}')\hat{\phi}({\mathbf{x}})
+\mbox{h.c.},\\
\hat{U}_3&=&\int \int d^3{\mathbf{x}} d^3{\mathbf{x}}' \hat{\phi}^\dagger({\mathbf{x}}) \hat{\Psi}^\dagger({\mathbf{x}}')
         U(|{\mathbf{x}}-{\mathbf{x}}'|)
	\hat{\Psi}({\mathbf{x}}')\hat{\phi}({\mathbf{x}}),\\
\hat{V}_1&=&\frac{1}{2}n_{0}^{2}\int \int d^3{\mathbf{x}} d^3{\mathbf{x}}' 
         V(|{\mathbf{x}}-{\mathbf{x}}'|),\\
\hat{V}_2&=& n_{0}\sqrt{n_0}\int \int d^3{\mathbf{x}} d^3{\mathbf{x}}' 
         V(|{\mathbf{x}}-{\mathbf{x}}'|)\hat{\phi}({\mathbf{x}})+\mbox{h.c.},\\
\hat{V}_3&=&\frac{1}{2}n_0\int \int d^3{\mathbf{x}} d^3{\mathbf{x}}' 
         V(|{\mathbf{x}}-{\mathbf{x}}'|)\hat{\phi}({\mathbf{x}}') 
\hat{\phi}({\mathbf{x}})+\mbox{h.c.},\\
\hat{V}_4&=& n_0\int \int d^3{\mathbf{x}} d^3{\mathbf{x}}'  
\hat{\phi}^{\dagger}({\mathbf{x}})
         V(|{\mathbf{x}}-{\mathbf{x}}'|)\hat{\phi}({\mathbf{x}}'),\\
\hat{V}_5&=& n_0\int \int d^3{\mathbf{x}} d^3{\mathbf{x}}'  
\hat{\phi}^{\dagger}({\mathbf{x}})
         V(|{\mathbf{x}}-{\mathbf{x}}'|)\hat{\phi}({\mathbf{x}}),\\
\hat{V}_6&=& \sqrt{n_0}\int \int d^3{\mathbf{x}} d^3{\mathbf{x}}'\hat{\phi}^{\dagger}({\mathbf{x}})
         V(|{\mathbf{x}}-{\mathbf{x}}'|)
         \hat{\phi}({\mathbf{x}}')\hat{\phi}({\mathbf{x}})
	 +\mbox{h.c.},\\
\hat{V}_7&=&\frac{1}{2}\int \int d^3{\mathbf{x}} d^3{\mathbf{x}}' \hat{\phi}^\dagger({\mathbf{x}}) \hat{\phi}^\dagger({\mathbf{x}}')
         V(|{\mathbf{x}}-{\mathbf{x}}'|)\hat{\phi}({\mathbf{x}}')\hat{\phi}({\mathbf{x}}).
\end{eqnarray}

\section{Systematic perturbation theory with Green's functions}
\label{systematic}
\subsection{Green's functions: definitions}
\label{green}
The boson ($B$) and fermion ($F$) Green's functions for the boson-fermion system
are defined as
\begin{eqnarray}
iG_{\mbox{\scriptsize B}}({\mathbf{x}},t,{\mathbf{x}}',t')&=&\langle 
\xi|T[\hat{\Phi}({\mathbf{x}},t)
\hat{\Phi}^\dagger({\mathbf{x}}',t')]|\xi\rangle,
\label{GFB}\\
iG_{\mbox{\scriptsize F}}({\mathbf{x}},t,{\mathbf{x}}',t')&=&\langle 
\xi|T[\hat{\Psi}({\mathbf{x}},t)
\hat{\Psi}^\dagger({\mathbf{x}}',t')]|\xi\rangle,	  
\label{GFF}
\end{eqnarray}
where the time argument in $\hat{\Phi}({\mathbf{x}},t)$ and 
$\hat{\Psi}({\mathbf{x}},t)$ means they evolve according to Heisenberg's equations of motion,
$T$ denotes the time ordered product, and $|\xi\rangle$ is the ground state of $\hat{K}$ (we
similarly define $|\xi_0\rangle$ to be the ground state of $\hat{K}_0$).
We use the Bogoliubov replacement to write
%(cf. pp. 204ff of Ref.\ \cite{FetWal71}):
\begin{equation}
iG_{\mbox{\scriptsize B}}({\mathbf{x}},t,{\mathbf{x}}',t')=n_0+iG_{\mbox{\scriptsize B}}'({\mathbf{x}},t,{\mathbf{x}}',t'),
\end{equation}
where 
\begin{equation}
iG'_{\mbox{\scriptsize B}}({\mathbf{x}},t,{\mathbf{x}}',t')=
  \langle \xi|T[\hat{\phi}({\mathbf{x}},t) 
  \hat{\phi}^\dagger({\mathbf{x}}',t')]|\xi\rangle
\end{equation}
is the propagator for the non-condensate bosons.

\subsection{Perturbative expansion}
The Green's functions can be evaluated in
perturbation theory
\cite{FetWal71}, 
where $\hat{W}$ is the perturbation to $\hat{K}_0$. Thus
\begin{eqnarray}
iG'_{\mbox{\scriptsize B}}({\mathbf{x}},t,{\mathbf{x}}',t')&=&
\frac{\sum_{n=0}^{\infty}i\tilde{G}_{\mbox{\scriptsize B}}^{(n)}({\mathbf{x}},t,{\mathbf{x}}',t')}
{\sum_{n=0}^{\infty}\langle \xi_0|S^{(n)}|\xi_0\rangle },
\label{GFexp}\\
iG_{\mbox{\scriptsize F}}({\mathbf{x}},t,{\mathbf{x}}',t')&=&
\frac{\sum_{n=0}^{\infty}i\tilde{G}_{\mbox{\scriptsize F}}^{(n)}({\mathbf{x}},t,{\mathbf{x}}',t')}
{\sum_{n=0}^{\infty}\langle \xi_0|S^{(n)}|\xi_0\rangle },
\label{GBexp}
\end{eqnarray}
where
\begin{eqnarray}
i\tilde{G}_{\mbox{\scriptsize B}}^{(n)}({\mathbf{x}},t,{\mathbf{x}}',t')&=&%\nonumber\\&&
\langle\xi_0|
T[S^{(n)}\tilde{\phi}({\mathbf{x}},t)\tilde{\phi}^\dagger({\mathbf{x}}',t')]
|\xi_0\rangle,
\label{Bnum}
\\
i\tilde{G}_{\mbox{\scriptsize F}}^{(n)}({\mathbf{x}},t,{\mathbf{x}}',t')&=&%\nonumber\\&&
\langle \xi_0|
T[S^{(n)}\tilde{\Psi}({\mathbf{x}},t)\tilde{\Psi}^\dagger({\mathbf{x}}',t')]
|\xi_0\rangle,
\label{Fnum}
\\
S^{(n)}
&=&
\label{denum}
\frac{1}{n!}\left(\frac{-i}{\hbar}\right)^n\int dt_1\ldots\int dt_n
T[\tilde{W}(t_1)\ldots \tilde{W}(t_n)].
\label{Smat}
\end{eqnarray}
Operators with a tilde are defined to be in the interaction picture, i.e.\ 
$\tilde{O}(t)=\exp(i\hat{K}_0t/\hbar)\hat{O}\exp(-i\hat{K}_0t/\hbar)$.
In the limit of a non-interacting system ($\hat{W}\rightarrow 0$) the
Green's functions reduce to the zeroth order terms in the expansions, so that
\begin{eqnarray}
iG_{\mbox{\scriptsize B}}^{0}({\mathbf{x}},t,{\mathbf{x}}',t')&=&
i\tilde{G}_{\mbox{\scriptsize B}}^{(0)}({\mathbf{x}},t,{\mathbf{x}}',t')
=\langle 
\xi_{0}|T[\tilde{\phi}({\mathbf{x}},t)
\tilde{\phi}^\dagger({\mathbf{x}}',t')]|\xi_{0}\rangle,
\label{GFBn}\\
iG_{\mbox{\scriptsize F}}^{0}({\mathbf{x}},t,{\mathbf{x}}',t')&=&
i\tilde{G}_{\mbox{\scriptsize F}}^{(0)}({\mathbf{x}},t,{\mathbf{x}}',t')
=\langle 
\xi_{0}|T[\tilde{\Psi}({\mathbf{x}},t)
\tilde{\Psi}^\dagger({\mathbf{x}}',t')]|\xi_{0}\rangle.	  
\label{GFFn}
\end{eqnarray}

\subsection{Evaluation of terms using Wick's theorem}
\label{contraction}
Equations (\ref{Bnum}), (\ref{Fnum}), and (\ref{denum}) can be evaluated by Wick's theorem, which
states that 
the vacuum (non-interacting ground-state)
expectation values of time ordered products of operators can be expressed 
as 
the sum of all products of contractions of pairs of operators in the time-ordered
product
\cite{Wic50}.
The contraction of two operators is defined as
\begin{equation}
\tilde{O}(t)^{(i)}\tilde{P}(t')^{(i)}=T[\tilde{O}(t)\tilde{P}(t')] \, -
\, :\tilde{O}(t)\tilde{P}(t'): \, ,
\end{equation}
where $:\tilde{O}(t)\tilde{P}(t'):$ is the normal ordered product. 
In particular,
\begin{eqnarray}
\tilde{\phi}({\mathbf{x}},t)^{(i)}\tilde{\phi}^\dagger({\mathbf{x}}',t')^{(i)}
&=&\tilde{\phi}^\dagger({\mathbf{x}}',t')^{(i)}\tilde{\phi}({\mathbf{x}},t)^{(i)}
=iG_{\mbox{\scriptsize B}}^0({\mathbf{x}},t,{\mathbf{x}}',t'),
\label{bosecontr}\\
\tilde{\Psi}({\mathbf{x}},t)^{(i)}\tilde{\Psi}^\dagger({\mathbf{x}}',t')^{(i)}
&=&-\tilde{\Psi}^\dagger({\mathbf{x}}',t')^{(i)}\tilde{\Psi}({\mathbf{x}},t)^{(i)}
=iG_{\mbox{\scriptsize F}}^0({\mathbf{x}},t,{\mathbf{x}}',t'),
\label{allcontr}
\end{eqnarray}
and all other contractions of pairs of operators 
$\in \{\tilde{\phi}({\mathbf{x}},t), \tilde{\phi}^\dagger({\mathbf{x}}',t'), 
\tilde{\Psi}({\mathbf{x}}'',t''),\tilde{\Psi}^\dagger({\mathbf{x}}''',t''')\}$ 
vanish (see also Appendix \ref{vacuum}).
Substituting Eqs.\ 
(\ref{bosecontr}) and (\ref{allcontr}) 
into Eqs.\ (\ref{Bnum}), (\ref{Fnum}), and (\ref{denum}), the first
order terms can be determined to be:
\begin{eqnarray}
i\tilde{G}_{\mbox{\scriptsize B}}^{(1)}(x^\mu,y^\mu)
&=&
\frac{-i}{\hbar}\int\int d^4x_{1}^\mu d^4y_{1}^\mu 
\biggl\{
U(x_1^\mu-y_1^\mu)
\nonumber\\&&\times
\left[-n_0iG^0_{\mbox{\scriptsize F}}(y_1^\mu,y_1^\mu)iG^0_{\mbox{\scriptsize B}}(x^\mu,y^\mu)
\right.\nonumber\\&&\left.
-iG^0_{\mbox{\scriptsize F}}(y_1^\mu,y_1^\mu)iG^0_{\mbox{\scriptsize B}}(x^\mu,x_1^\mu)iG^0_{\mbox{\scriptsize B}}(x_1^\mu,y^\mu)\right]
\nonumber\\&&
+V(x_1^\mu-y_1^\mu)
\biggl[
\frac{n_0^2}{2}iG^0_{\mbox{\scriptsize B}}(x^\mu,y^\mu)
\nonumber\\&&
+n_0iG^0_{\mbox{\scriptsize B}}(x^\mu,x_1^\mu)iG^0_{\mbox{\scriptsize B}}(x_1^\mu,y^\mu)
\nonumber\\&&
+n_0iG^0_{\mbox{\scriptsize B}}(x^\mu,x_1^\mu)iG^0_{\mbox{\scriptsize B}}(x^\mu,x_2^\mu)
\biggr]\biggr\},
\label{NB1}
\\
i\tilde{G}_{\mbox{\scriptsize F}}^{(1)}(x^\mu,y^\mu)
&=&
\frac{-i}{\hbar}\int\int d^4x_{1}^\mu d^4y_{1}^\mu 
\biggl\{
U(x_1^\mu-y_1^\mu)
\nonumber\\&&\times
[-n_0 iG^0_{\mbox{\scriptsize F}}(y_1^\mu,y_1^\mu) iG^0_{\mbox{\scriptsize F}}(x^\mu,y^\mu)
\nonumber\\&&
-n_0 iG^0_{\mbox{\scriptsize F}}(x^\mu,y_1^\mu) iG^0_{\mbox{\scriptsize F}}(y_1^\mu,y^\mu)]
\nonumber\\ 
&&+
V(x_1^\mu-y_1^\mu)
\frac{n_0^2}{2}
iG_{\mbox{\scriptsize F}}(x^\mu,y^\mu)
\biggr\}
\label{NF1}
\\
S^{(1)}
&=&
\frac{-i}{\hbar}\int\int d^4x_{1}^\mu d^4y_{1}^\mu 
\biggl\{
U(x_1^\mu-y_1^\mu)
\nonumber\\ &&\times
[-i n_0 G^0_{\mbox{\scriptsize F}}(y_1^\mu,y_1^\mu)]+
V(x_1^\mu-y_1^\mu)
\frac{n_0^2}{2}
\biggr\},
\label{D1}
\end{eqnarray}
where we have  used the more compact four-vector notation 
[$x^\mu=(t,{\mathbf{x}})$], and defined 
$U(x^\mu-y^\mu)=U({\mathbf x}-{\mathbf y})\delta(x^0-y^0)$ and
$V(x^\mu-y^\mu)=V({\mathbf x}-{\mathbf y})\delta(x^0-y^0)$. 
Note that
$G^0_{\mbox{\scriptsize B}}(t,{\mathbf x},t, {\mathbf y})=0$ (i.e. there are no boson loops
at zero temperature).

Higher order terms may be similarly evaluated,
and will similarly be expressed in terms of integrals over products of
noninteracting Green's functions, condensate factors $n_{0}$, 
and interaction terms.
% \label{feynman}
We represent these graphically (see Fig.\ \ref{figdef}):
straight lines for fermions, wiggly lines for non-condensate bosons, dashed lines
for condensate bosons, and zigzag lines for interaction terms (whether it is a
boson-boson or boson-fermion interaction is clearly determined by the kinds of
particle lines attached to the vertices of the interaction line):

\begin{figure}[h]
\begin{center}
\epsfxsize=8.5cm
\epsfbox{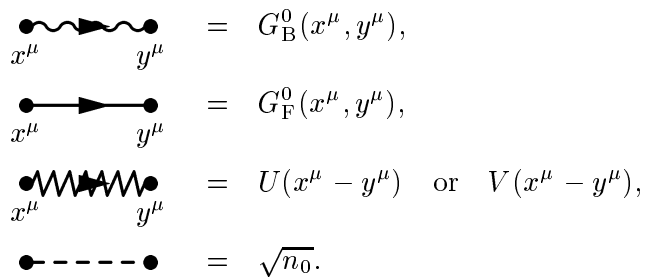}
\end{center}
\caption{Definition of the diagram lines}
\label{figdef}
\end{figure} 

As is usual \cite{AbrGor63,FetWal71,mod}, 
all disconnected graphs in the numerator can be
factorized out by the denominator, so that
\begin{eqnarray}
G'_{\mbox{\scriptsize B}}(x^\mu,y^\mu)&=&
\sum_{n=0}^{\infty}
\tilde{G}_{\mbox{\scriptsize B}}^{(n)}(x^\mu,y^\mu)_{\mbox{\scriptsize 
connected}} \; , \\
G_{\mbox{\scriptsize F}}(x^\mu,y^\mu)&=&
\sum_{n=0}^{\infty}
\tilde{G}_{\mbox{\scriptsize F}}^{(n)}(x^\mu,y^\mu)_{\mbox{\scriptsize 
connected}} \; .
\end{eqnarray}
Noting that each connected graph
essentially appears
$n!$ times, with simple permutations on the labeling,
when composing such graphs  
we integrate over all internal variables and affix a factor 
of $(i/\hbar)^n(-1)^F(-i)^C$, where $n$ is the 
number of interaction lines, $F$ is the number of closed fermion loops, and 
$C$ is the number of dashed boson lines.

\subsection{Feynman rules}
For homogeneous systems it is convenient to
Fourier transform to energy-momentum space, so that: 
\begin{eqnarray}
G^0_{\mbox{\scriptsize B}}(p^\mu)&=&\frac{1}{p^0-\hbar p^2/2m_{\mbox{\scriptsize B}}+\mu_{B}/\hbar+i\nu},
\label{GF0BM}
\\
G^0_{\mbox{\scriptsize F}}(p^\mu)&=&\frac{1}{p^0-\hbar p^2/2m_{\mbox{\scriptsize F}}+i \mbox{sgn}(p-k_{\mbox{\scriptsize F}})\nu},
\label{GF0FM}
\end{eqnarray}
where $\mbox{sgn}(k)=1$ for $k\geq 0$ and %$\mbox{sgn}(x)
$=-1$ for $k<0$ (we write $p$ for $|\mathbf{p}|$). 
The appropriate Feynman rules for the boson (fermion) Green's function in this 
representation are then:

1) Draw all topologically distinct connected diagrams 
with 
one outgoing external
   wiggly boson (fermion) line and one incoming external wiggly boson (fermion) 
   line, no external
   fermion (boson) lines and no internal dashed boson lines,
$n$ zigzag interaction lines, each of which is attached at one vertex to
	an incoming and an outgoing boson line (either wiggly or dashed), and at
	the other vertex either to an incoming and an outgoing boson line, or to an
	incoming and an outgoing (not necessarily distinct) fermion line.
	Each vertex must be attached to exactly one zigzag interaction line.

2) All wiggly boson lines must run into the same direction and there are no 
   closed boson loops.

3) Each dashed boson line corresponds to a factor of $\sqrt{n_0}$, each wiggly boson line 
   to a factor of $G^0_{\mbox{\scriptsize B}}(k^\mu)$, each fermion line to a factor of $G^0_{\mbox{\scriptsize F}}(k^\mu)$,
   each boson-fermion interaction line to a factor of $U(k^\mu)=U({\mathbf{k}})$, and
   each boson-boson interaction line to a factor of $V(k^\mu)=V({\mathbf{k}})$.

4) Assign a direction to each interaction line; associate a directed four-momentum with each line
   and conserve four-momentum at each vertex. Each dashed boson line 
   carries four-momentum $0$ and each wiggly boson line has four-momentum $\neq 0$.

5) Integrate over the $n$ independent four-momenta.

6) Affix a factor of $(i/\hbar)^{n} (2\pi)^{-4(n)}(-1)^F(-i)^C$, where $F$ is the 
   number of closed
   fermion loops and $C$ is the number of dashed boson lines.\\

\section{Determination of the boson-fermion $T$-matrix and self-energies 
in ladder approximation}
\label{determination}
\subsection{The Hugenholtz-Pines theorem}
According to the Hugenholtz-Pines theorem \cite{HugPin59,HPTcomment}, the bosonic chemical potential 
$\mu_{\mbox{\scriptsize B}}$,
defined as
\begin{equation}
\frac{\partial E/V}{\partial n_{\mbox{\scriptsize B}}}=\mu_{\mbox{\scriptsize B}},
\label{chempotdef}
\end{equation}
is given by
\begin{eqnarray}
\mu_{\mbox{\scriptsize B}} &=&\hbar\Sigma_{\mbox{\scriptsize B}}(0)-\hbar\Sigma_{12}(0),
\label{hugpinthm}
\end{eqnarray}
where $\Sigma_{\mbox{\scriptsize B}}(0)$ and $\Sigma_{12}(0)$ are the proper 
self-energies for the bosons due to their interaction with both bosons and fermions, 
evaluated at $p^{\mu}=0$ (in what follows we call them the bosonic 
self-energies). 
The self-energies are in general related to the Green's
functions by the Dyson equations. 
The Dyson equation for the bosons is given by:
\begin{eqnarray}
&& \left( \begin{array}{cc} G_{\mbox{\scriptsize B}}'(p^\mu) & G_{12}(-p^\mu) \\
                            G_{21}(p^\mu) & G_{\mbox{\scriptsize B}}'(-p^\mu) \end{array} \right)
  = \left( \begin{array}{cc} G_{\mbox{\scriptsize B}}^0(p^\mu) & 0 \\
                            0 & G_{\mbox{\scriptsize B}}^0(-p^\mu)
                            \end{array} \right) \nonumber \\ 
&& \nonumber \\
&& + \left( \begin{array}{cc} G_{\mbox{\scriptsize B}}^0(p^\mu) & 0 \\
                            0 & G_{\mbox{\scriptsize B}}^0(-p^\mu) \end{array} \right)
%			    \nonumber \\ && \times
   \left( \begin{array}{cc} \Sigma_{\mbox{\scriptsize B}}(p^\mu) & \Sigma_{12}(p^\mu) \\
                            \Sigma_{21}(p^\mu) & \Sigma_{\mbox{\scriptsize B}}(-p^\mu) 
			    \end{array} \right)
%			    \nonumber\\&&\times
   \left( \begin{array}{cc} G_{\mbox{\scriptsize B}}'(p^\mu) & G_{12}(p^\mu) \\
       G_{21}(p^\mu) & G_{\mbox{\scriptsize B}}'(-p^\mu) \end{array} 
\right) \; ,
%			    \nonumber \\
\label{DEBM}
\end{eqnarray}
where have introduced the anomalous boson Green's functions
$G_{12}(p^\mu)$ and $G_{21}(p^\mu)$
(defined as the Fourier transforms of
$G_{12}(x^\mu,y^\mu)=
\langle \xi|T[\hat{\phi}(x^\mu)\hat{\phi}(y^\mu)]|\xi\rangle$ and
$G_{21}(x^\mu,y^\mu)=
\langle {\bf
G}|[\hat{\phi}^\dagger(x^\mu)\hat{\phi}^\dagger(y^\mu)]|\xi\rangle$,
respectively).
The Dyson equation for the fermions takes the much simpler 
scalar form:
\begin{equation}
G_{\mbox{\scriptsize F}}(p^\mu)=G_{\mbox{\scriptsize
F}}^0(p^\mu)+G_{\mbox{\scriptsize F}}^0(p^\mu)\Sigma_{\mbox{\scriptsize
F}}(p^\mu)G_{\mbox{\scriptsize F}}(p^\mu) \; ,
\label{DEFS}
\end{equation}
where $\Sigma_{\mbox{\scriptsize F}}(p^\mu)$ is the proper self-energy for the fermions due to the 
interaction with the bosons (the fermionic self-energy). 

\subsection{The self-energies in ladder approximation}
\label{ladder}
\begin{figure}[h]
\begin{center}
\epsfxsize=8.5cm
\epsfbox{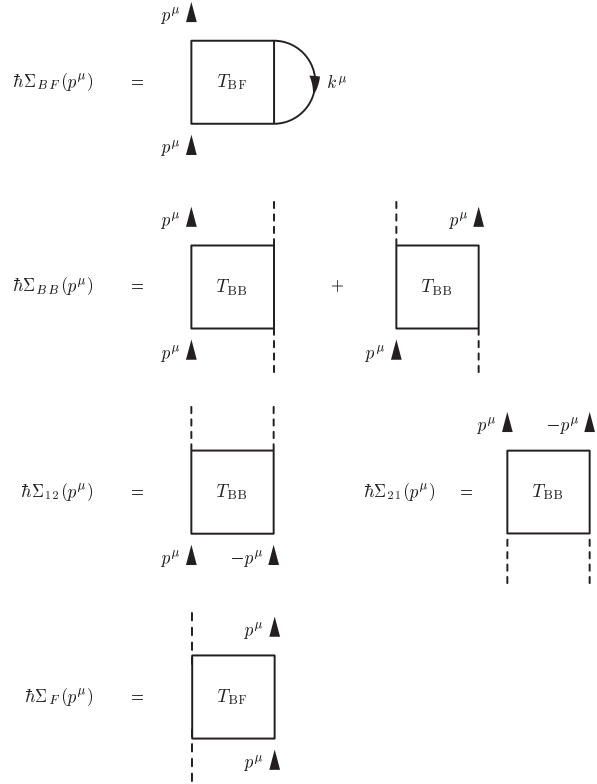}
\end{center}
\caption{The self-energies in ladder approximation, expressed in terms of the $T$-matrices.}
\label{selfenergies}
\end{figure}

\begin{figure}[h]
\begin{center}
\epsfxsize=15.7cm
\epsfbox{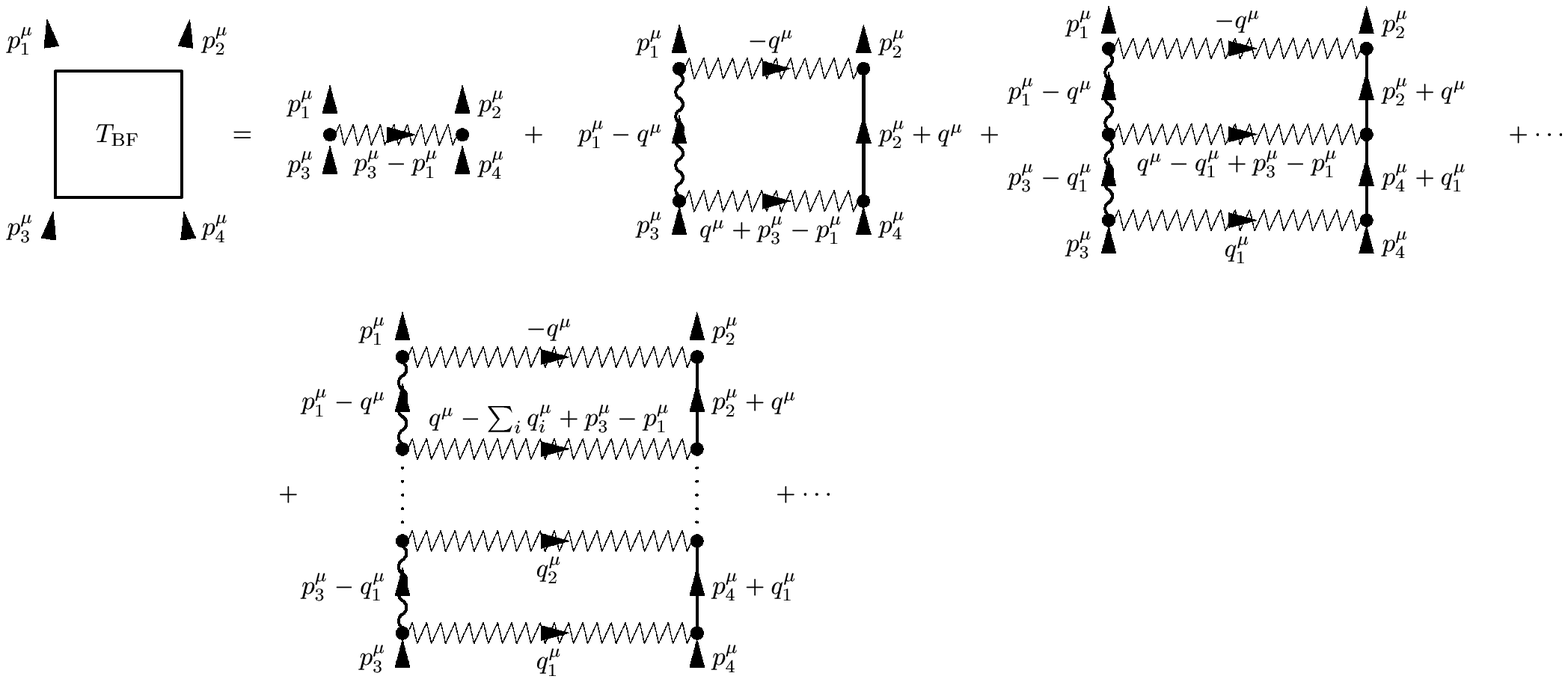}
\end{center}
\caption{The boson-fermion $T$-matrix.}
\label{tmatrix0}
\end{figure}

As we are considering a dilute system,
in terms of Feynman diagrams  only diagrams with interaction lines between 
two systems of connected 
propagators %(one for the fermions, the other for the bosons) 
are important \cite{AbrGor63,FetWal71} (ladder approximation).
This is expressed in terms of the boson-fermion and boson-boson $T$-matrices
in Fig.\ \ref{selfenergies}, where the boson-fermion $T$-matrix 
$T_{\mbox{\scriptsize BF}}$ in ladder approximation
is defined in Fig.\ \ref{tmatrix0}, the
boson-boson $T$-matrix $T_{\mbox{\scriptsize BB}}$ (also in ladder approximation) 
is well known from studies of dilute pure Bose systems, and the
normal (diagonal) bosonic proper self-energy is given by  
\begin{equation}
\Sigma_{\mbox{\scriptsize B}}(p^{\mu}) = 
\Sigma_{\mbox{\scriptsize BF}}(p^{\mu}) + 
\Sigma_{\mbox{\scriptsize BB}}(p^{\mu}).
\label{properself}
\end{equation}
The proper self-energies can thus be determined by adding the proper self-energies of a system of
interacting bosons to those of a hypothetical 
mixed boson-fermion system where there are boson-fermion interactions
only \cite{feschbach}. This result arises from our use of the ladder approximation, and is not in
general true (there also exist, for example, 
inseparable three-legged ``ladders''
consisting of a boson-boson and a boson-fermion ladder
joined by a common boson leg, but these clearly involve
three-particle processes). 
For such a hypothetical mixed system, the only self-energies we need to 
consider and to evaluate
are $\Sigma_{\mbox{\scriptsize BF}}(p^\mu)$ and $\Sigma_{\mbox{\scriptsize F}}(p^\mu)$, which can
be written algebraically as:
\begin{eqnarray}
\hbar\Sigma_{\mbox{\scriptsize BF}}(p^\mu)&=&-\frac{i}{(2\pi)^{4}}\int
d^4k^\mu T_{\mbox{\scriptsize
BF}}(p^\mu,k^\mu,p^\mu,k^\mu)G^0_{\mbox{\scriptsize F}}(k^\mu), 
\label{sigmaB} \\
\hbar\Sigma_{\mbox{\scriptsize F}}(p^\mu)&=&T_{\mbox{\scriptsize BF}}(0,p^\mu,0,p^\mu)n_0.
\label{sigmaF}
\end{eqnarray}

\subsection{Bethe-Salpeter equation for $T_{\mbox{\scriptsize BF}}$}
\begin{figure}[h]
\begin{center}
\epsfxsize=12cm
\epsfbox{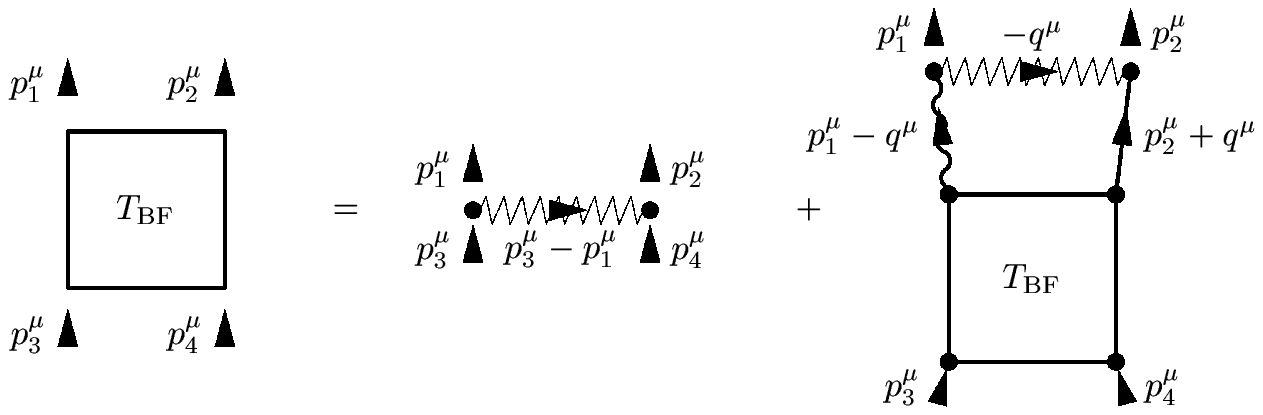}
\end{center}
\caption{The integral equation for $T_{\mbox{\scriptsize BF}}$.}
\label{tmatrix}
\end{figure}
The boson-fermion $T$-matrix $T_{\mbox{\scriptsize BF}}$
can also be represented recursively, as shown in Fig.\ \ref{tmatrix}.
If we now transform to center-of-mass coordinates,
\begin{eqnarray}
P^\mu&=&p_1^\mu+p_2^\mu=p_3^\mu+p_4^\mu,\nonumber\\
k_1^\mu&=&(p_1^\mu-p_2^\mu)/2,\nonumber\\ 
k_2^\mu&=&(p_3^\mu-p_4^\mu)/2,
\end{eqnarray}
the algebraic form of the equation represented in Fig.\ \ref{tmatrix} reads:
\begin{eqnarray}
T_{\mbox{\scriptsize
BF}}(k_1^\mu,k_2^\mu,P^\mu)&=&U({\mathbf{k}}_1-{\mathbf{k}}_2) 
+\frac{i}{\hbar(2\pi)^4}\int 
d^3{\mathbf{k}}U({\mathbf{k}}_1-{\mathbf{k}}) \nonumber \\
&&\times\int dk^0 G^0_{\mbox{\scriptsize B}}(P^\mu/2+k^\mu)
G^0_{\mbox{\scriptsize F}}(P^\mu/2-k^\mu)T_{\mbox{\scriptsize
BF}}(k^\mu,k_2^\mu,P^\mu) \; .
\label{BSECM}
\end{eqnarray}
This is a kind of Bethe-Salpeter integral equation, 
which we will now solve recursively for low momenta,
stopping at order $a_{\mbox{\scriptsize BF}}^{2}$ . 
As the interactions are instantaneous, the only frequency dependence
in $T_{\mbox{\scriptsize BF}}(k_1^\mu,k_2^\mu,P^\mu)$ is in $P^0$ \cite{AbrGor63,FetWal71}. 
Thus, a contour integration over $k^{0}$ in Eq.~(\ref{BSECM})
yields:
\begin{eqnarray}
&& T_{\mbox{\scriptsize BF}}({\mathbf{k}}_1,{\mathbf{k}}_2,P^\mu) =
U({\mathbf{k}}_1-{\mathbf{k}}_2)
\nonumber \\ &&
+ \frac{1}{(2\pi)^3}\int d^3{\mathbf{k}}
%U({\mathbf{k}}_1-{\mathbf{k}})
\frac{
U({\mathbf{k}}_1-{\mathbf{k}})
T_{\mbox{\scriptsize BF}}({\mathbf{k}},{\mathbf{k}}_2,P^\mu)\theta(|{\mathbf{P}}/2-{\mathbf{k}}|-k_{\mbox{\scriptsize F}})}
{\hbar P^0-\hbar^2({\mathbf{P}}/2+{\mathbf{k}})^2/2m_{\mbox{\scriptsize B}}-
\hbar^2({\mathbf{P}}/2-{\mathbf{k}})^2/2m_{\mbox{\scriptsize F}}+\mu+i\nu}.
%T_{\mbox{\scriptsize BF}}({\mathbf{k}},{\mathbf{k}}_2,P^\mu) \; .
\label{BSECM2}
\end{eqnarray}
We now express Eq.~(\ref{BSECM2}) in terms of
the free scattering amplitude $f({\mathbf{k}}_1,{\mathbf{k}}_2)$, 
first by formally inversion (see \cite{AbrGor63}):
\begin{eqnarray}
\frac{2\pi\hbar^2}{m}f({\mathbf{k}}_1,{\mathbf{k}}_2)&=&U({\mathbf{k}}_2-{\mathbf{k}}_1)
+\frac{1}{(2\pi)^3}\int d^3{\mathbf{k}}
\frac{U({\mathbf{k}}_2-{\mathbf{k}})2\pi\hbar^2 f({\mathbf{k}}_1,{\mathbf{k}})/m}
{\hbar^2{\mathbf{k}}_1^2/2m-\hbar^2{\mathbf{k}}^2/2m +i\nu} \; ,
\label{scatteringampl}
\end{eqnarray}
and then by exploiting the resulting expression to
rewrite Eq.\ (\ref{BSECM2}) as:
\begin{eqnarray}
T_{\mbox{\scriptsize BF}}({\mathbf{k}}_1,{\mathbf{k}}_2,P^\mu)&=&\frac{2\pi\hbar^2}{m}f({\mathbf{k}}_2,{\mathbf{k}}_1)
+\frac{1}{(2\pi)^3}\int d^3{\mathbf{k}}
\frac{2\pi\hbar^2}{m}f({\mathbf{k}},{\mathbf{k}}_1)T_{\mbox{\scriptsize BF}}({\mathbf{k}},{\mathbf{k}}_2,P^\mu)
\nonumber\\
&&\times\left[\frac{\theta(|{\mathbf{P}}/2-{\mathbf{k}}|-k_{\mbox{\scriptsize F}})}
{\hbar P^0-\hbar^2({\mathbf{P}}/2+{\mathbf{k}})^2/2m_{\mbox{\scriptsize B}}-\hbar^2
({\mathbf{P}}/2-{\mathbf{k}})^2/2m_{\mbox{\scriptsize F}}+\mu+i\nu}
\right.\nonumber\\&&\left.
-\frac{1}{\hbar^2{\mathbf{k}}_1^2/2m-\hbar^2{\mathbf{k}}^2/2m+i\nu}
\right] \, .
\label{BSECM3}
\end{eqnarray}
For low momenta the vacuum scattering 
amplitude $f({\mathbf{k}}_1,{\mathbf{k}}_2)$ can be 
expanded to second order in the
scattering length $a_{\mbox{\scriptsize BF}}$ (see \cite{FetWal71}):
\begin{equation}
f({\mathbf{k}}_1,{\mathbf{k}}_2)\approx 
a_{\mbox{\scriptsize BF}}-ia_{\mbox{\scriptsize BF}}^2k, 
\label{fexp}
\end{equation}
where $k = k_{1} = k_{2} \rightarrow 0$.
We insert this into Eq.\ (\ref{BSECM3}), iteratively substituting
Eq.\ (\ref{BSECM3}) into itself, and consistently keeping terms
up to quadratic order in $a_{\mbox{\scriptsize BF}}$ only. This produces
\begin{eqnarray}
T_{\mbox{\scriptsize BF}}({\mathbf{k}}_1,{\mathbf{k}}_2,P^\mu)&\approx&
\frac{2\pi\hbar^2}{m}[a_{\mbox{\scriptsize BF}}-ia_{\mbox{\scriptsize BF}}^2k_1]
+\frac{4\pi^2\hbar^4 a_{\mbox{\scriptsize BF}}^2}{(2\pi)^3 m ^2}\nonumber\\
&&\times\int d^3{\mathbf{k}}\left[\frac{\theta(|{\mathbf{P}}/2-{\mathbf{k}}|-k_{\mbox{\scriptsize F}})}
{\hbar P^0-\hbar^2({\mathbf{P}}/2+{\mathbf{k}})^2
/2m_{\mbox{\scriptsize B}}-\hbar^2({\mathbf{P}}/2-{\mathbf{k}})^2/2m_{\mbox{\scriptsize F}}+\mu+i\nu}
\right.\nonumber\\&&\left.
-\frac{1}{\hbar^2{\mathbf{k}}_1^2/2m-\hbar^2{\mathbf{k}}^2/2m+i\nu}\right],
\label{BSECM5}
\end{eqnarray} 
the renormalized second order expansion of the boson-fermion $T$-matrix. 
The integral can be evaluated 
(see App.\ \ref{integral})
to give
\begin{eqnarray}
&& T_{\mbox{\scriptsize BF}}({\mathbf{k}}_1,{\mathbf{k}}_2,P^\mu) \approx
\frac{2\pi\hbar^2}{m}a_{\mbox{\scriptsize BF}}+\frac{2\hbar^2 
a_{\mbox{\scriptsize BF}}^2 k_{\mbox{\scriptsize F}}}{m}\nonumber\\
&& \nonumber \\
&&+\frac{a_{\mbox{\scriptsize BF}}^2\hbar^2}{2m^2}\left(
\frac{m_{\mbox{\scriptsize B}} k_{\mbox{\scriptsize F}}^2}{P}-\frac{m^2P}
{m_{\mbox{\scriptsize B}}}-2m\sqrt{D}
-\frac{m_{\mbox{\scriptsize B}} D}{P}\right)
\ln\frac{k_{\mbox{\scriptsize F}}+
mP/m_{\mbox{\scriptsize B}}+\sqrt{D}+
i\nu/2a_{\mbox{\scriptsize BF}}
\sqrt{D}}{k_{\mbox{\scriptsize F}}
-mP/m_{\mbox{\scriptsize B}}-\sqrt{D}
-i\nu/2a_{\mbox{\scriptsize BF}}\sqrt{D}}\nonumber\\
&& \nonumber \\
&&-\frac{a_{\mbox{\scriptsize BF}}^2\hbar^2}{2m^2}\left(\frac{m_{\mbox{\scriptsize B}} k_{\mbox{\scriptsize F}}^2}{P}-\frac{m^2P}
{m_{\mbox{\scriptsize B}}}+2m\sqrt{D}-\frac{m_{\mbox{\scriptsize B}} D}{P}\right)
\ln\frac{k_{\mbox{\scriptsize F}}-mP/m_{\mbox{\scriptsize
B}}+\sqrt{D}+i\nu/2a_{\mbox{\scriptsize
BF}}\sqrt{D}}{k_{\mbox{\scriptsize F}}+mP/m_{\mbox{\scriptsize
B}}-\sqrt{D}-i\nu/2a_{\mbox{\scriptsize BF}}\sqrt{D}} \; , 
\label{TMat}
\end{eqnarray}
where
\begin{equation}
D=-\frac{m}{m_{\mbox{\scriptsize B}}+m_{\mbox{\scriptsize
F}}}P^2+\frac{2mP^0}{\hbar}+\frac{2m\mu}{\hbar^2} \; .
\label{Dabbr}
\end{equation}

\section{Physical quantities}
\label{results}
\subsection{Bosonic chemical potential}

Substituting Eq.~(\ref{GF0FM}) into  Eq.\ (\ref{sigmaB}) the equation  for $\Sigma_{\mbox{\scriptsize BF}}(p^{\mu})$ 
can be rewritten as
\begin{eqnarray}
\hbar\Sigma_{\mbox{\scriptsize BF}}(p^\mu)&=&-\frac{i}{(2\pi)^{4}}
\int d^4q^\mu
%\times 
\frac{T_{\mbox{\scriptsize BF}}[({\mathbf{p}}-{\mathbf{q}})/2,({\mathbf{p}}-{\mathbf{q}})/2,p^\mu+q^\mu]}
{q^0-\hbar{\mathbf{q}}^2/2m_{\mbox{\scriptsize F}}+i
\mbox{sgn}(q-k_{\mbox{\scriptsize F}})\nu} \; .
\label{SigmaBexpr}
\end{eqnarray}
To evaluate this, we substitute Eq.\ (\ref{BSECM5}) into Eq.\ (\ref{SigmaBexpr}),
and first carry out  the frequency integral.
As the pole in the complex $q^0$-plane of 
the integrand in Eq.\ (\ref{BSECM5}) is below the real axis,
in order to get a non vanishing result
the pole of $[q^0-\hbar q^2/2m_{\mbox{\scriptsize F}}-i\mbox{sgn}(q-k_{\mbox{\scriptsize F}})\nu]^{-1}$ 
must be above the real axis ($q<k_{\mbox{\scriptsize F}}$). 
The frequency integral is thus readily solved by contour integration. 
The ${\mathbf{k}}$ integration in 
(\ref{BSECM5}) is then very similar to that leading to Eq.\ (\ref{TMat}). 
The resulting expression for $\hbar\Sigma_{\mbox{\scriptsize BF}}(p^\mu)$ is then:
\begin{equation}
\hbar\Sigma_{\mbox{\scriptsize BF}}(p^\mu)
=\frac{1}{(2\pi)^{3}}\int d^3{\mathbf{q}}\theta(k_{\mbox{\scriptsize F}}-q)
T_{\mbox{\scriptsize BF}}\left[\frac{{\mathbf{p}}-{\mathbf{q}}}{2},
\frac{{\mathbf{p}}-{\mathbf{q}}}{2},
\left(p^0+\frac{\hbar{\mathbf{q}}^2}{2m_{\mbox{\scriptsize F}}},
{\mathbf{p}}+{\mathbf{q}}
\right)\right] .
\label{Sigmabfint}
\end{equation}

We wish to similarly solve this integral to second order in $a_{\mbox{\scriptsize BF}}$.
In Eq.\ (\ref{TMat}), all terms which depend on $D$ have
a pre-factor $a_{\mbox{\scriptsize BF}}^2$. Thus, in order to get a result 
for Eq.~(\ref{Sigmabfint})
that is correct to second order in $a_{\mbox{\scriptsize BF}}$,
it is sufficient to use the zeroth order expression for $D$.
Specializing to the case where $p^\mu=0$ 
%(necessary for implementation of the Hugenholtz-Pines theorem) 
this can be written as
\begin{equation}
D^0=\frac{m_{\mbox{\scriptsize B}}^2}{(m_{\mbox{\scriptsize B}}+m_{\mbox{\scriptsize F}})^2}q^2.
\end{equation}
We now substitute $D^0$ for $D$ in Eq.\ (\ref{TMat}), and, after a straightforward (if lengthy) 
integration over ${\mathbf{q}}$, arrive at
\begin{eqnarray}
\hbar\Sigma_{\mbox{\scriptsize BF}}(0)&=&
\frac{2\pi\hbar^2a_{\mbox{\scriptsize BF}}}{m}n_{\mbox{\scriptsize F}}
\left[
1 + \frac{a_{\mbox{\scriptsize BF}}k_{\mbox{\scriptsize F}}}{\pi}f(\delta)
\right],
\label{SigmaB2}
\end{eqnarray} 
where 
\begin{equation}
f(\delta) = 
1-\frac{3+\delta}{4\delta}
+\frac{3(1+\delta)^{2}(1-\delta)}{8\delta^{2}}\ln
\frac{1+\delta}{1-\delta} \, ,
\label{FM}
\end{equation}
$\delta=(m_{\mbox{\scriptsize B}}-m_{\mbox{\scriptsize F}})/(m_{\mbox{\scriptsize B}}+m_{\mbox{\scriptsize F}})$, and we have used
$
(2\pi)^{-4}\int d^4k^\mu G_{\mbox{\scriptsize F}}(k^\mu)=G_{\mbox{\scriptsize F}}(x^\mu,x^\mu)=i n_{\mbox{\scriptsize F}}
$. Note that in this integration we need only consider the real part of the boson-fermion
$T$-matrix, as 
within the range of the integration 
the imaginary part is zero (see App.\ \ref{integral}). 
The necessary expression for the $T$-matrix is then just given by Eq.\
(\ref{TMat}),
where we take the absolute values of the arguments of the logarithms and set $\nu=0$.
From the Hugenholtz-Pines theorem
[Eq.\ (\ref{hugpinthm})]:
\begin{eqnarray}
\mu_{\mbox{\scriptsize B}} 
&=&\hbar\Sigma_{\mbox{\scriptsize BF}}(0)+\hbar\Sigma_{\mbox{\scriptsize BB}}(0)-\hbar\Sigma_{12}(0).
\end{eqnarray}
Thus, using the expression for $\Sigma_{\mbox{\scriptsize BF}}(0)$ in Eq.\ (\ref{SigmaB2}), 
and the results from \cite{Bel58} 
for $\Sigma_{\mbox{\scriptsize BB}}(0)$ and $\Sigma_{12}(0)$ 
(neglecting corrections of the order of the boson gas parameter), 
\begin{equation}
\mu_{\mbox{\scriptsize B}}=
\frac{4\pi\hbar^2 
a_{\mbox{\scriptsize BB}}}{m_{\mbox{\scriptsize B}}}
n_{\mbox{\scriptsize B}}
+\frac{2\pi\hbar^2a_{\mbox{\scriptsize BF}}}{m}n_{\mbox{\scriptsize F}}
\left[1+\frac{a_{\mbox{\scriptsize BF}}k_{\mbox{\scriptsize F}}}{\pi}f(\delta)
\right].
\label{muB}
\end{equation} 
This is exactly equivalent to adding $\hbar\Sigma_{\mbox{\scriptsize BF}}(0)$ to the standard mean-field result for the 
bosonic chemical potential for a pure, self-interacting bosonic system.

\subsection{Ground state energy density}
\label{groundE}
To obtain the ground state energy we simply integrate Eq.~(\ref{chempotdef}):
\begin{equation}
\frac{E}{V}=\int_0^{n_{\mbox{\scriptsize B}}}\mu(n_{\mbox{\scriptsize B}}´)dn_{\mbox{\scriptsize B}}´+C(n_{\mbox{\scriptsize F}}),
\label{intsol}
\end{equation}
where $C(n_{\mbox{\scriptsize F}})$ is a quantity  that can depend on the fermion density $n_{\mbox{\scriptsize F}}$ only. 
%(and not on $n_{\mbox{\scriptsize B}}$).
Considering the limit $a_{\mbox{\scriptsize BF}} \to 0$, we see that $C(n_{\mbox{\scriptsize F}})$ can only be the kinetic energy for 
free fermions (the Fermi energy density $\epsilon_{\mbox{\scriptsize
F}}$)\cite{FetWal71}, that is:
\begin{equation}
C(n_{\mbox{\scriptsize F}})=\epsilon_{\mbox{\scriptsize F}}=\frac{3}{5}\frac{\hbar^2k_{\mbox{\scriptsize F}}^2}{2m_{\mbox{\scriptsize F}}}n_{\mbox{\scriptsize F}}.
\end{equation}
Substituting this and Eq.\ (\ref{muB}) into Eq.\ (\ref{intsol}), and then integrating, gives, finally:
\begin{equation}
\frac{E}{V}=\epsilon_{\mbox{\scriptsize F}}+\epsilon_{\mbox{\scriptsize B}}
+\frac{2\pi\hbar^2a_{\mbox{\scriptsize BF}}}{m}n_{\mbox{\scriptsize F}}n_{\mbox{\scriptsize B}}
\left[1+\frac{a_{\mbox{\scriptsize BF}}k_{\mbox{\scriptsize F}}}{\pi}f(\delta)
\right],
\label{E0}
\end{equation}
where $\epsilon_{\mbox{\scriptsize B}}=2\pi\hbar^2a_{\mbox{\scriptsize BB}}
n_{\mbox{\scriptsize B}}^{2}/m_{\mbox{\scriptsize B}}$ is the 
boson mean-field energy density.
Eq.\ (\ref{E0}) is the main result of this paper, being the desired 
extension of the mean-field result Eq.\ (\ref{mfe}).

\begin{figure}[h]
\begin{center}
\epsfxsize=8.5cm
\epsfbox{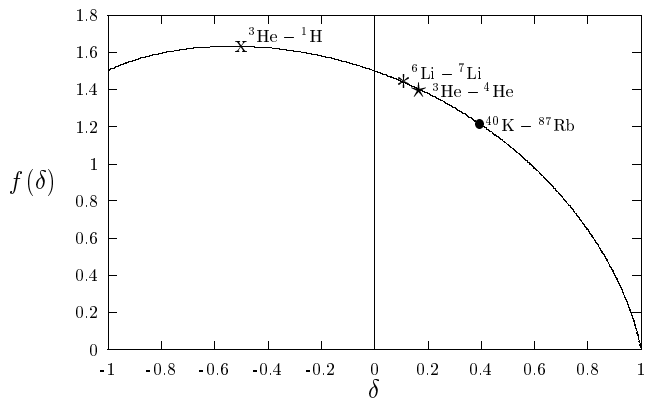}
\end{center}
\caption{Plot of $f(\delta)$, 
where $\delta = (m_{\mbox{\scriptsize B}} - 
m_{\mbox{\scriptsize F}})/(m_{\mbox{\scriptsize B}} + 
m_{\mbox{\scriptsize F}})$, 
proportional to
the correction to second order in 
$\alpha=k_{\mbox{\scriptsize F}}a_{\mbox{\scriptsize BF}}$ 
to the energy-density
functional [Eq.\ (\ref{E0scaled})]. The relevant values of $f(\delta)$ for 
mixtures of
$^{3}$He and $^{1}$H,
$^{6}$Li and $^{7}$Li, 
$^{3}$He and $^{4}$He, and 
$^{40}$K and $^{87}$Rb are
indicated.}
\label{mplot}
\end{figure}
It is illuminating to describe Eq.~(\ref{E0}) in terms of the dimensionless
gas parameters and the dimensionless ratio of the boson and fermion 
densities:
\begin{eqnarray}
\alpha&=&a_{\mbox{\scriptsize BF}}k_{\mbox{\scriptsize F}},\\
\beta&=&\sqrt{n_{\mbox{\scriptsize B}}a_{\mbox{\scriptsize BB}}^3},\\
\eta&=&\frac{n_{\mbox{\scriptsize B}}}{n_{\mbox{\scriptsize F}}},
\end{eqnarray}
so that
\begin{equation}
\frac{E}{V}=\epsilon_{\mbox{\scriptsize F}}\left(1+\frac{20\pi\eta}{1+\delta}
\Biggl\{(1-\delta)\left(\frac{\eta\beta}{6\pi^2}\right)^{2/3}
\right. \left.
+\frac{\alpha}{3\pi^2}
\left[1+\frac{\alpha}{\pi}f(\delta)\right]\Biggr\}\right),
\label{E0scaled}
\end{equation}
The corrective term to second order in $\alpha$ is proportional 
to the rather
complicated function $f(\delta)$, defined in Eq.\ (\ref{FM}), 
of the relative mass ratio 
$\delta$; the value of this function
will thus vary considerably depending on the masses of
the atomic species 
used in any given experiment. In Fig.\ \ref{mplot} the values
for mixtures of $^{6}$Li and $^{7}$Li, and $^{40}$K and $^{87}$Rb,
corresponding to real
experimental configurations currently under investigation, are plotted, 
as well as that for a mixture of
$^{3}$He and $^{4}$He, also a likely candidate for future investigation in
ultra-cold dilute gas experiments. 
The value for a hypothetical mixture of $^{3}$He and $^{1}$H is also 
shown, as it is almost exactly the maximum possible.
The function $f(\delta)$ is always positive in the total range
$[-1,1]$ of variations of $\delta$.
Note that in the limit 
$m_{\mbox{\scriptsize B}}/m_{\mbox{\scriptsize F}}\to\infty$, one
has
$\delta\to 1$ and $f(\delta)\to 0$. 
Thus the second-order correction to the boson-fermion interaction
energy and the total boson-boson 
interaction energy disappear.
This is because if the bosons are infinitely 
massive (compared to a fixed, finite fermion
mass), then it is impossible for them to be
scattered out of the condensate, and 
only the boson-fermion mean field interaction remains, since all the
bosons can be treated as the (condensate) mean field.
In the opposite limit 
of $m_{\mbox{\scriptsize B}}/m_{\mbox{\scriptsize F}}\to 0$, 
the situation is different, because of the 
Pauli exclusion principle. 
%for the fermions, and the total 
%interaction energy density consequently diverges. 

\begin{figure}[h]
\begin{center}
\epsfxsize=14.5cm
\epsfbox{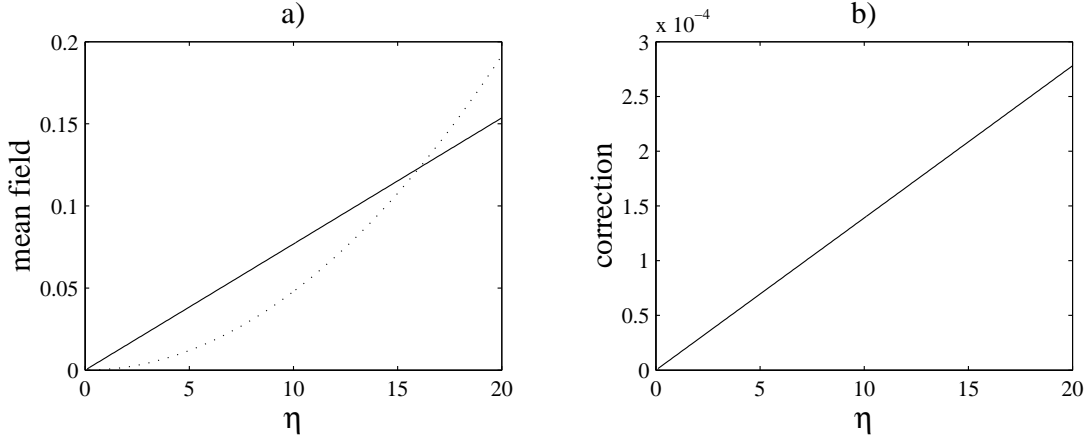}
\end{center}
\caption{ 
Boson-boson (solid line) and 
boson-fermion (dotted line) mean field contributions (a) 
and second-order corrections (b) to the
ground-state energy-density of a $^{6}$Li, $^{7}$Li mixture, in 
units of $\epsilon_{\mbox{\scriptsize F}}$, 
to the ground-state energy-density in units of 
$\epsilon_{\mbox{\scriptsize F}}$. Parameters are 
$n_{\mbox{\scriptsize F}}^{1/3}a_{\mbox{\scriptsize BF}}=0.001$ 
and $n_{\mbox{\scriptsize B}}^{1/3}a_{\mbox{\scriptsize BB}} = 
0.000135\eta^{1/3}$, 
where the density ratio $\eta$ is the running independent 
variable.}
\label{Li}
\end{figure}
\begin{figure}[h]
\begin{center}
\epsfxsize=14.5cm
\epsfbox{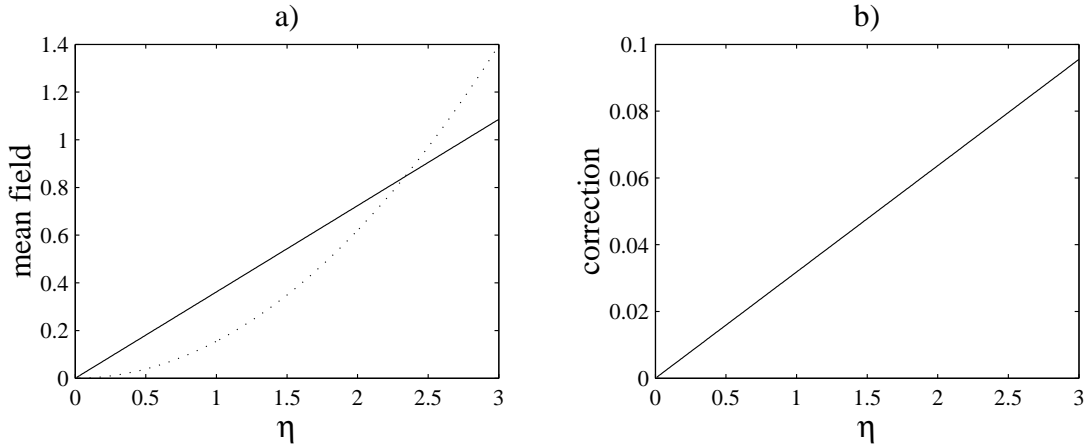}
\end{center}
\caption{Boson-boson (solid line) 
and boson-fermion (dotted line) mean field contributions (a) 
and second-order corrections (b) to the
ground-state energy-density of a $^{3}$He, 
$^{4}$He mixture, in units of $\epsilon_{\mbox{\scriptsize F}}$, 
to the ground-state energy-density 
in units of $\epsilon_{\mbox{\scriptsize F}}$. 
Parameters are 
$n_{\mbox{\scriptsize F}}^{1/3}a_{\mbox{\scriptsize BF}}=0.05$ 
and $n_{\mbox{\scriptsize B}}^{1/3}a_{\mbox{\scriptsize BB}} 
= 0.05\eta^{1/3}$, where the density ratio $\eta$ is the 
running independent variable.}
\label{He}
\end{figure}

In Fig.\ \ref{Li} we compare the mean field contributions (a) 
and second-order correction (b) to the
energy functional 
for a $^{6}$Li, $^{7}$Li mixture, for a range of values of $\eta$.
The plots correspond to a situation 
where the scattering 
lengths $a_{\mbox{\scriptsize BB}}
=0.2$nm, $a_{\mbox{\scriptsize BF}}=2.7$nm  and the fermion density
$n_{\mbox{\scriptsize F}}=5.1\times 10^{10}$cm$^{-3}$
are fixed, and compatible 
with the experiments described in Ref.\ \cite{SchKha01}, while
the boson density is varied. 
Note that for any reasonable boson density, the boson gas parameter $\beta$ is indeed very small compared to
$\alpha$.  
%(the plot for the beyond mean-field corrections is extended 
%out to $\eta = 300$ in order to show their crossover).
In Fig.\ \ref{He} we do the same for a 
$^{3}$He, $^{4}$He mixture. 
In this case the inter-species scattering length is unknown,
however we conjecture it to be of the 
same order of magnitude as the boson-boson scattering length. The
plots correspond to a 
situation where $a_{\mbox{\scriptsize BB}}=a_{\mbox{\scriptsize BF}}=16$nm, 
and $n_{\mbox{\scriptsize F}}=3.1\times 10^{13}$cm$^{-3}$. 
These values are compatible with current experiments on
metastable triplet $^{4}$He condensates \cite{RobSir01,PerLeo01}, 
and are particularly
interesting in that the beyond 
mean-field corrections are quite large (of the order of
10\%). The true significance of the boson-fermion interaction 
energy correction will of
course depend on the actual value of the interspecies scattering length.
We notice that if the latter will turn out to be of about
one order of magnitude larger than in the pure fermionic and 
bosonic cases (this is for instance what happens for lithium 
mixtures), then the effect of the correction can be as large
as $50$\% of the mean-field prediction. Then, of course, also 
corrections proportional to the boson gas parameter have to be taken into account.

\subsection{Other physical quantities, Bose condensate
depletion, and 
induced fermion-fermion interaction}
From Eq.\ (\ref{E0}) we can readily determine
the chemical potential for the 
fermions $\mu_{\mbox{\scriptsize F}}$, defined as
\begin{equation}
\mu_{\mbox{\scriptsize F}}
=\left(\frac{\partial E/V}{\partial 
n_{\mbox{\scriptsize F}}}\right)_{N_{\mbox{\scriptsize B}},V},
\end{equation}
to be
\begin{eqnarray}
\mu_{\mbox{\scriptsize F}}&=&
\frac{\hbar^2k_{\mbox{\scriptsize F}}^2}{2m_{\mbox{\scriptsize F}}}
+\frac{2\pi\hbar^2a_{\mbox{\scriptsize BF}}}{m}n_{\mbox{\scriptsize B}}
\left[
1 + \frac{4a_{\mbox{\scriptsize BF}}k_{\mbox{\scriptsize F}}}{3\pi}f(\delta)
\right].
\end{eqnarray}
The pressure reads
\begin{eqnarray}
P&=&-\left(\frac{\partial 
E}{\partial V}\right)_{N_{\mbox{\scriptsize B}},N_{\mbox{\scriptsize F}}}
\nonumber \\
&& \nonumber \\
&=&\frac{2}{5}\frac{\hbar^2k_{\mbox{\scriptsize F}}^2}{2m_{\mbox{\scriptsize 
F}}}n_{\mbox{\scriptsize F}}
+\frac{2\pi n_{\mbox{\scriptsize 
B}}^2a_{\mbox{\scriptsize BB}}\hbar^2}{m_{\mbox{\scriptsize B}}}
%\nonumber \\&&
+\frac{2\pi\hbar^2a_{\mbox{\scriptsize 
BF}}}{m}n_{\mbox{\scriptsize F}}n_{\mbox{\scriptsize B}}
\left[1+\frac{4a_{\mbox{\scriptsize 
BF}}k_{\mbox{\scriptsize F}}}{3\pi}f(\delta)
\right].
\end{eqnarray}
We then obtain for the bosonic and fermionic compressibilities,
respectively, for the bosons:
\begin{eqnarray}
&& c_{\mbox{\scriptsize B}}^2 = \frac{1}{m_{\mbox{\scriptsize 
B}}}\left(\frac{\partial P}{\partial n_{\mbox{\scriptsize 
B}}}\right)_{N_{\mbox{\scriptsize F}},V} \nonumber \\
&& \nonumber \\
&& =
\frac{4\pi n_{\mbox{\scriptsize B}}^2a_{\mbox{\scriptsize 
BB}}\hbar^2}{m^2_{\mbox{\scriptsize B}}}
+\frac{2\pi\hbar^2a_{\mbox{\scriptsize 
BF}}}{m_{\mbox{\scriptsize B}}m}n_{\mbox{\scriptsize F}}
\left[1+\frac{4a_{\mbox{\scriptsize 
BF}}k_{\mbox{\scriptsize F}}}{3\pi}f(\delta)
\right],
\end{eqnarray}
and for the fermions:
\begin{eqnarray}
c_{\mbox{\scriptsize F}}^2&=&\frac{1}{m_{\mbox{\scriptsize 
F}}}\left(\frac{\partial P}{\partial n_{\mbox{\scriptsize 
F}}}\right)_{N_{\mbox{\scriptsize B}},V} \nonumber \\
&& \nonumber \\
&=&\frac{2}{3}\frac{\hbar^2k_{\mbox{\scriptsize 
F}}^2}{2m^2_{\mbox{\scriptsize F}}}
+\frac{2\pi\hbar^2a_{\mbox{\scriptsize 
BF}}}{m_{\mbox{\scriptsize F}}m}n_{\mbox{\scriptsize B}}
\left[1+\frac{16a_{\mbox{\scriptsize 
BF}}k_{\mbox{\scriptsize F}}}{9\pi}f(\delta)\right].
\end{eqnarray}
We notice that the possible instabilities induced by the
mean-field boson-fermion interaction term in the case of
a negative value of $a_{\mbox{\scriptsize BF}}$ are contrasted
by the beyond mean-field correction, since the latter
is always positive.

Concerning the structure of the Bose condensate fraction, 
besides the known depletion due to  
the boson-boson interaction, we expect in
principle a further contribution to depletion due to the
interaction of the bosons with the fermions.
The depletion is computed in a standard way by integrating
the boson propagator for the non condensed particles
$G_{\mbox{\scriptsize B}}'(p^\mu)$ over four-momentum.
To obtain the boson propagator we have to solve the
Dyson equation (\ref{DEBM}) for $G_{\mbox{\scriptsize B}}'(p^\mu)$.
This yields:
\begin{eqnarray}
G_{\mbox{\scriptsize B}}'(p^\mu) &=& 
%\nonumber \\&& 
\left[p^0 + \frac{\hbar p^2}{2m_{\mbox{\scriptsize B}}} +
\Sigma_{\mbox{\scriptsize B}}(-p^{\mu}) - \Sigma_{\mbox{\scriptsize B}}(0)
+ \Sigma_{12}(0)\right]
\nonumber\\&&\times
\left\{\left[ p^0 - \frac{\Sigma_{\mbox{\scriptsize B}}(p^{\mu})
- \Sigma_{\mbox{\scriptsize B}}(-p^{\mu})}{2} \right]^2
\right.\nonumber\\&&\left.
-\left[ \frac{\hbar p^2}{2m_{\mbox{\scriptsize B}}}
- \Sigma_{\mbox{\scriptsize B}}(0) + \Sigma_{12}(0)
+ \frac{\Sigma_{\mbox{\scriptsize B}}(p^{\mu})
+ \Sigma_{\mbox{\scriptsize B}}(-p^{\mu})}{2} \right]^2 
+ \Sigma_{12}^{2}(p^{\mu})\right\}^{-1}, 
%\nonumber \\&& 
\label{bosonpropagator}
\end{eqnarray}

\noindent where we have made use of the Hugenholtz-Pines
relation. The total diagonal bosonic self-energy 
$\Sigma_{\mbox{\scriptsize B}}(p^{\mu})$ picks up
a boson-boson and a boson-fermion contribution (see
Eq.\ (\ref{properself})). It can be easily
checked that to first order in
$a_{\mbox{\scriptsize BB}}$ and
in $a_{\mbox{\scriptsize BF}}$
the total diagonal bosonic self-energy 
does not depend
on the four-momentum. Therefore the diagonal self-energy
terms in the boson propagator Eq.\ (\ref{bosonpropagator}) 
cancel, $G_{\mbox{\scriptsize B}}'(p^\mu)$ is
independent of $a_{\mbox{\scriptsize BF}}$,
and there is no depletion
of the Bose condensate due to the fermions
to this order, since
the contribution of the fermions to the off-diagonal 
self-energies vanishes anyway in ladder approximation.
The situation will be different to next order in
$a_{\mbox{\scriptsize BF}}$, as in this case the
total diagonal bosonic self-energy will depend on the
four-momentum. However, the calculation of 
$\Sigma_{\mbox{\scriptsize B}}(p^{\mu})$ to second
order in the boson-fermion scattering length 
at non zero four-momentum involves the evaluation of
integrals that cannot be carried out
analytically in a straightforward way.
In conclusion, in the present situation, we will
consider the depletion due to the bosons only, which is well known
\cite{FetWal71,AbrGor63}:
\begin{equation}
n_{\mbox{\scriptsize B}}-n_0=\frac{8}{3}\sqrt{\frac{n_{\mbox{\scriptsize B}}a_{\mbox{\scriptsize B}}^3}{\pi}}
                             n_{\mbox{\scriptsize B}} \, .
\label{depletion}
\end{equation}

We now turn to the discussion of the fermion-fermion 
interaction induced by the presence of the bosons.
Subtracting the bosonic contribution from the energy density, we get
\begin{equation}
\frac{E}{V}-\epsilon_{\mbox{\scriptsize B}} =
\epsilon_{\mbox{\scriptsize F}}\left[1+\frac{20\eta}{3\pi(1+\delta)}\alpha+
\right. \left.
\frac{20 \eta}{3\pi(1+\delta)}\frac{f(\delta)}{\pi}\alpha^2\right].
\end{equation}
This describes the first
three terms of a power expansion in $\alpha$, of exactly 
the same form as that of an imperfect
(unpolarized) Fermi gas, though clearly with 
different coefficients. There is thus, as expected, 
an induced fermion-fermion interaction, which can now be
computed by exploiting the expressions that we have
derived for both the bosonic and the fermionic
chemical potentials. This will yield a modification
of the known induced fermion-fermion interaction 
previously discussed in mean field approximation \cite{VivPet00}. 
%The following discussion is very analogous to this case. 
The expression for the
induced interaction at zero energy-momentum transfer 
$U_{\mbox{\scriptsize ind}}(q^{\mu} = 0)$
is 
%\cite{VivPet00}
:
\begin{equation}
U_{\mbox{\scriptsize ind}}(q^{\mu} = 0) = \left.  
\frac{\partial \mu_{\mbox{\scriptsize F}}}{\partial 
n_{\mbox{\scriptsize F}}}
\right|_{\mu_{\mbox{\scriptsize B}}} 
- \frac{\partial \mu_{\mbox{\scriptsize F}}}{\partial 
n_{\mbox{\scriptsize F}}}
= -\left(\frac{\partial \mu_{\mbox{\scriptsize F}}}{\partial 
n_{\mbox{\scriptsize B}}}\right)^{2}
\frac{\partial n_{\mbox{\scriptsize B}}}{\partial 
\mu_{\mbox{\scriptsize B}}} \, , 
\end{equation}

\noindent which in our case reads:
\begin{equation}
U_{\mbox{\scriptsize ind}}(q^{\mu} = 0) =
-\frac{4\pi\hbar^{2}(1-\delta^{2})\eta^{1/3}\alpha^{2}}
{m_{\mbox{\scriptsize F}}k_{\mbox{\scriptsize F}}(6\pi^{2})^{1/3}\beta^{2/3}}
\left[1+4f(\delta)\alpha /3\pi\right].
\label{Uind0}
\end{equation}

\begin{figure}[h]
\begin{center}
\epsfxsize=4.5cm
\epsfbox{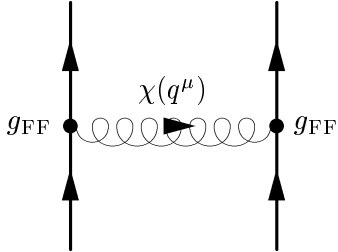}
\end{center}
\caption{Effective fermion-fermion interaction due
to exchange of boson density fluctuations.}
\label{Uind}
\end{figure}

The extension to finite momentum transfer
is achieved by introducing the boson density-density
response function $\chi(q^{\mu})$, and is represented
diagrammatically in Fig.\ \ref{Uind}, where two fermions
interact by exchanging a boson density fluctuation wave. 
This is the simplest diagram by which two fermions can
interact via the exchange of a bosonic excitation.

The density-density response function $\chi(q^{\mu})$ 
is independent of $a_{\mbox{\scriptsize BF}}$
to first order. This can be easily verified
following the same line of reasoning described
previously in the analysis of the bosonic
propagator.
Thus its expression is the same as in the pure bosonic
case (to this order):
\begin{equation}
\chi(q^{\mu}) = 
\frac{\hbar n_0 q^2 /m_{\mbox{\scriptsize B}}}{(q^0)^2
- (\hbar q^2 /2m_{\mbox{\scriptsize B}})^2 -
4 \pi \hbar^3 n_0 
a_{\mbox{\scriptsize BB}}q^2/m_{\mbox{\scriptsize B}}^2} \, .
\end{equation}

The vertex $g_{\mbox{\scriptsize FF}}$ can be determined
by considering the limiting expression for $\chi(q^{\mu})$
as $q^\mu \rightarrow 0$. In this case the expression
for the diagram given in Fig.\ \ref{Uind} must reduce
to the expression given in Eq.\ (\ref{Uind0}).
It follows that:
\begin{equation}
g_{\mbox{\scriptsize FF}}=\frac{2\sqrt{\pi}\hbar^2a_{\mbox{\scriptsize BF}}}{m}
\left[1+4f(\delta)a_{\mbox{\scriptsize BF}}k_{\mbox{\scriptsize F}}/3\pi\right].
\end{equation}
Then the induced interaction potential in the static case ($q^0=0$) reads:
\begin{equation}
U_{\mbox{\scriptsize ind}}(0,{\vec q})
=-\frac{4\pi\hbar^4a^2_{\mbox{\scriptsize BF}}\left(1+
4f(\delta)a_{\mbox{\scriptsize BF}}k_{\mbox{\scriptsize F}}/3\pi\right)^2}{m^2}
\frac{n_0}{
\hbar q^2 /4m_{\mbox{\scriptsize B}} +4 \pi \hbar^2 n_0 
a_{\mbox{\scriptsize BB}}/m_{\mbox{\scriptsize B}}} \, .
\end{equation}
In real space this is:
\begin{eqnarray}
U_{\mbox{\scriptsize ind}}(0,{\vec q})=
\frac{4\hbar^2m_{\mbox{\scriptsize B}}n_0a^2_{\mbox{\scriptsize BF}}
\left(1+4f(\delta)a_{\mbox{\scriptsize BF}}k_{\mbox{\scriptsize F}}/3\pi\right)^2}{m^2}
\frac{e^{\sqrt{2}r/\xi}}{r} \, ,
\end{eqnarray}
where 
\begin{equation}
\xi^2=\frac{1}{8\pi n_0a_{\mbox{\scriptsize B}}} \, .
\end{equation}
We observe that, compared to the mean-field
result \cite{VivPet00}, while there are 
quantitative modifications in the pre-factors, 
there is no qualitative change in the form of 
the induced interaction,
i.e. we still have an attractive Yukawa potential.
Modifications in the analytic form of the induced
fermion-fermion interaction potential will appear
only once second-order effects in the 
boson-fermion scattering length
$a_{\mbox{\scriptsize BF}}$ to the depletion
of the Bose condensate are included.

\section{Discussion and Outlook}
\label{discussion}
In summary, we have determined ground-state 
properties of a homogeneous system of bosons mixed with
spin-polarized fermions at zero temperature. 
We have calculated the boson-fermion $T$-matrix and 
the corresponding self-energies. 
Then we have shown how to incorporate the 
effects of the boson-boson interaction
and derived some relevant physical 
quantities of the system, in particular
the ground-state energy. The importance
of the beyond mean-field corrections has
been discussed in several different instances
of experimental interest. For mixtures of
bosonic and fermionic Helium we have shown
that the beyond mean-field terms may yield 
significant corrections (up to $50$\% of
the mean-field result).
We have provided partial results also on
two very significant physical quantities,
namely the Bose condensate fraction and
the induced fermion-fermion interaction.
To provide more quantitative predictions
for these quantities, as well as for the
BCS transition temperature, we will need
to compute in detail the corrections to
second order in the boson-fermion scattering
length. Results of this analysis, which goes
beyond the scope of the present paper, will
appear in a forthcoming work, together with
a detailed numerical analysis of the conditions
for stability and for phase separation.
Collective modes, effective fermion mass, and
excitation spectra fully evaluated to second
order in the boson-fermion scattering length
will be discussed as well.

Extensions of the formalism developed in this
paper can be made in different directions. 
First, one can consider unpolarized spin-$1/2$ fermions.
Calculations are very similar to 
the present situation with the main difference that one has to include the
effects of the direct interactions of fermions with different spins. 
This would correspond to having a third
scattering length $a_{\mbox{\scriptsize FF}}$. 
As in the previous case, 
we expect that in the two-particle scattering approximation
the energy contribution for a 
pure fermion system of spin-$1/2$ 
fermions is simply added to Eq.\ (\ref{E0}) in this case
(and of course, the Fermi 
momentum has to be modified appropriately). 

The formalism can be also extended to consider
finite temperature. 
As in the case of pure bosons 
we expect considerable difficulties near the critical
temperature. 
Well below that temperature, 
however, we expect no major
complications and the calculations will
be similar to the present 
case, except that boson loops will
have to be taken into account and frequency integrals
will have to be replaced by Matsubara frequency sums. 

A very important and natural possible extension is the 
investigation of 
inhomogeneous, e.g. harmonically trapped, systems. 
This can be done by augmenting the existing mean field 
calculations via the correlations terms 
in local density approximation. To this end, the results 
obtained in the present work are needed. 
The method and the full numerical procedure will be described 
in a forthcoming paper.

Finally, higher-order corrections may be
in principle computed.
These higher-order terms will involve also the bosonic gas parameter
as well as three-particle correlations, 
and thus expansions like Eq.\ (\ref{E0})
will not reduce to sums of terms, 
where only one scattering length appears at a time, but 
will include also
terms which contain products of powers of
both scattering lengths. 
To even higher orders non-universal properties 
like the parameters describing the
shape of the interaction potentials will 
become important and will have to be taken 
properly into account, as it has been recently
done in the pure bosonic case \cite{BraNie,Braa01}.

\section*{Acknowledgements}
We are grateful to  
Gordon Baym for seminal comments
on an earlier draft of the present work. 
We aknowledge very useful discussions
with Sam Morgan and Stefano Giorgini,
as well as with Misha Baranov, Allan Griffin, Chris Pethick,
and Luciano Viverit.

A. P. Albus, S. A. Gardiner, and M. Wilkens
thank the DFG, BEC2000+, and the Alexander von Humboldt 
Foundation for financial support. F. Illuminati
thanks the INFM for financial support.

\begin{appendix}
\section{Normal ordered products and the vacuum state}
\label{vacuum}
If we expand the field operators in terms of momentum eigenstates we get:
\begin{eqnarray}
\hat{\Phi}({\mathbf{x}})&=&\frac{1}{\sqrt{V}}\sum_{{\mathbf{k}}} \hat{a}_{{\mathbf{k}}} e^{i{\mathbf{k}}\cdot{\mathbf{x}}}=
                     \frac{1}{\sqrt{V}}\hat{a}_0+\hat{\phi}({\mathbf{x}}),\\		     
\hat{\Psi}({\mathbf{x}})&=&\frac{1}{\sqrt{V}}\sum_{{\mathbf{k}}} \hat{b}_{{\mathbf{k}}} e^{i{\mathbf{k}}\cdot{\mathbf{x}}}=
       \hat{\psi_1}^\dagger({\mathbf{x}})+\hat{\psi_2}({\mathbf{x}}),
\end{eqnarray}
with 
\begin{eqnarray}
\hat{\phi}({\mathbf{x}})&=&\frac{1}{\sqrt{V}}\sum_{|{\mathbf{k}}|>0}\hat{a}_{{\mathbf{k}}} 
e^{i{\mathbf{k}}\cdot{\mathbf{x}}}, \\
\hat{\psi_1}({\mathbf{x}})&=&\frac{1}{\sqrt{V}}\sum_{|{\mathbf{k}}|\leq k_{\mbox{\scriptsize F}}} 
\hat{b}^\dagger_{{\mathbf{k}}} e^{-i{\mathbf{k}}\cdot{\mathbf{x}}} \; , \\
\hat{\psi_2}({\mathbf{x}})&=&\frac{1}{\sqrt{V}}\sum_{|{\mathbf{k}}|>k_{\mbox{\scriptsize F}}} 
\hat{b}_{{\mathbf{k}}} e^{i{\mathbf{k}}\cdot{\mathbf{x}}} \; .
\end{eqnarray} 
In terms of the bosonic and fermionic occupation number operators 
$\hat{N}_{\mbox{\scriptsize B}}({\mathbf{k}})=
\hat{a}^\dagger_{{\mathbf{k}}}\hat{a}_{{\mathbf{k}}}$ and 
$\hat{N}_{\mbox{\scriptsize F}}({\mathbf{k}})=\hat{b}^\dagger_{{\mathbf{k}}}\hat{b}_{{\mathbf{k}}}$ the
ground state of the non-interacting system $|\xi_{0}\rangle$ can be 
characterized by:
\begin{eqnarray}
\hat{N}_{\mbox{\scriptsize B}}(\vec{0})|\xi_{0}\rangle&=&N_{\mbox{\scriptsize B}}|\xi_{0}\rangle, \\
\hat{N}_{\mbox{\scriptsize B}}({\mathbf{k}})|\xi_{0}\rangle&=&0  \mbox{ for }|{\mathbf{k}}|>0, \\
\hat{N}_{\mbox{\scriptsize F}}({\mathbf{k}})|\xi_{0}\rangle&=& |\xi_{0}\rangle \mbox{ for }
|{\mathbf{k}}|\leq k_{\mbox{\scriptsize F}}, \\
\hat{N}_{\mbox{\scriptsize F}}({\mathbf{k}})|\xi_{0}\rangle&=&0 
\mbox{for} |{\mathbf{k}}|>k_{\mbox{\scriptsize F}} \, ,
\end{eqnarray}
where $N_{\mbox{\scriptsize B}}$ is the total number of bosons, which in this case coincides with
the number of zero-momentum bosons $N_{0}$ (Bose-Einstein condensate).
In occupation number representation we thus have 
\begin{equation}
|\xi_{0}\rangle=
|N_{\mbox{\scriptsize B}},0,0,\dots>_{\mbox{\scriptsize
|B}}\otimes|1,1,\ldots,1,1,0,0,\ldots>_{\mbox{\scriptsize F}} \; ,
\end{equation}
where the subscript B refers to the boson Hilbert space and the F to the 
fermion Hilbert space. The change from $1$ to $0$
in the fermion state happens at $k_{\mbox{\scriptsize F}}$.
Additionally,
\begin{equation}
\hat{\phi}({\mathbf{x}})|\xi_{0}\rangle=
\hat{\psi_1}({\mathbf{x}})|\xi_{0}\rangle=
\hat{\psi_2}({\mathbf{x}})|\xi_{0}\rangle=0.
\end{equation}
In this sense the ground state can be regarded as the vacuum state with 
respect to fermions excited above the Fermi sea, the
fermion holes below the Fermi sea, 
and the non-condensate bosons. 

The normal product is defined on pairs of creation and destruction operators:
\begin{eqnarray}
:\tilde{\phi}({\mathbf{x}},t)\tilde{\phi}^\dagger({\mathbf{x}}',t'):&=&\tilde{\phi}^\dagger({\mathbf{x}}',t')\tilde{\phi}({\mathbf{x}},t),
\nonumber\\
:\tilde{\psi}_j({\mathbf{x}},t)\tilde{\psi}_k^\dagger({\mathbf{x}}',t'):&=&-\tilde{\psi}_k^\dagger({\mathbf{x}}',t')\tilde{\psi}_j({\mathbf{x}},t),
\nonumber\\
:\tilde{\phi}({\mathbf{x}},t)\tilde{\psi}_j^\dagger({\mathbf{x}}',t'):&=&\tilde{\psi}_j^\dagger({\mathbf{x}}',t')\tilde{\phi}({\mathbf{x}},t),
\nonumber\\
:\tilde{\psi}_j({\mathbf{x}},t)\tilde{\phi}^\dagger({\mathbf{x}}',t'):&=&\tilde{\phi}^\dagger({\mathbf{x}}',t')\tilde{\psi}_j({\mathbf{x}},t),
\label{normalnormal}
\end{eqnarray}
for $j,k\in\{1,2\}$.
For all other pairs of creation and destruction operators the normal product is the same 
as the ordinary operator product. 
It can also be readily determined that
\begin{eqnarray}
[\tilde{\phi}({\mathbf{x}},t),\tilde{\phi}^\dagger(\vec{x'},t')]
&=&\langle \xi_0|\tilde{\phi}({\mathbf{x}},t)\tilde{\phi}^\dagger({\mathbf{x}}',t')|\xi_0\rangle,
\nonumber\\
\{\tilde{\psi}_{1}({\mathbf{x}}',t'),\tilde{\psi}^\dagger_{1}({\mathbf{x}},t)\}
&=&\langle \xi_0|\tilde{\Psi}^\dagger({\mathbf{x}}',t')\tilde{\Psi}({\mathbf{x}},t)|\xi_0\rangle,
\nonumber\\
\{\tilde{\psi}({\mathbf{x}},t),\tilde{\psi}^\dagger(\vec{x'},t')\}
&=&\langle \xi_0|\tilde{\Psi}({\mathbf{x}},t)\tilde{\Psi}^\dagger({\mathbf{x}}',t')|\xi_0\rangle,
\label{comanticom}
\end{eqnarray}
and all other (anti-)commutators are zero.
With Eqs.\ \ref{normalnormal} and \ref{comanticom}, the contractions 
%by distinguishing the  cases $t>t'$ and $t<t'$ and get 
of Eq.\ (\ref{allcontr}) can be readily evaluated.

\section{Evaluation of the $T$-Matrix and coupling constant renormalization}
\label{integral}
\subsection{The first integral ${\cal I}$}
We define
\begin{eqnarray}
{\cal I}=\int d^3{\mathbf{k}}\frac{\theta(|{\mathbf{P}}/2-{\mathbf{k}}|-k_{\mbox{\scriptsize F}})}
{\hbar P^0-\hbar^2({\mathbf{P}}/2+{\mathbf{k}})^2/2m_{\mbox{\scriptsize B}}-\hbar^2({\mathbf{P}}/2-{\mathbf{k}})^2/2m_{\mbox{\scriptsize F}}
+\mu+i\nu}.
\end{eqnarray}  
Transforming the integration variables to ${\mathbf{P}}/2-{\mathbf{k}}$ gives:
\begin{eqnarray}
{\cal I}=\int d^3{\mathbf{k}}\frac{\theta(|{\mathbf{k}}|-k_{\mbox{\scriptsize F}})}
{\hbar^2{\mathbf{k}}^2/2m-\hbar^2{\mathbf{P}}\cdot{\mathbf{k}}/m_{\mbox{\scriptsize B}}-
\hbar P^0+\hbar^2{\mathbf{P}}^2/2m_{\mbox{\scriptsize B}}-\mu-i\nu}.
\end{eqnarray} 
Setting 
$a=\hbar^2/2m$, $b=\hbar^2 P/m_{\mbox{\scriptsize B}}$ and 
$E=-\hbar P^0+\hbar^2 P^2/2m_{\mbox{\scriptsize B}}-\mu$ and transforming to spherical coordinates we get:
\begin{eqnarray}
{\cal I}&=&
2\pi \int^{k_c}_{k_{\mbox{\scriptsize F}}} dkk^2\int_0^\pi d\phi\sin\phi \frac{1}{ak^2-bk\cos\phi+E-i\nu}\nonumber \\
&=&
\frac{2\pi}{b} \int^{k_c}_{k_{\mbox{\scriptsize F}}} dkk\ln\frac{ak^2-bk+E-i\nu}{ak^2+bk+E-i\nu},
\end{eqnarray}  
where we will ultimately consider the limit $k_{c}\to \infty$.
Using 
$D=(b/2a)^2-E/a$
%=-\frac{m}{m_{\mbox{\scriptsize B}}+m_{\mbox{\scriptsize F}}}P^2+\frac{2mP^0}{\hbar}+\frac{2m\mu}{\hbar^2}$
we can approximate for small $\nu$ (if $D\neq0$; the case $D=0$ can be treated similarly and gives
the same answer as taking the limit $D\to 0$ at the very end):
\begin{eqnarray}
{\cal I}&=&
-\frac{2\pi m_{\mbox{\scriptsize B}}}{\hbar^2P} \int^{k_c}_{k_{\mbox{\scriptsize F}}} dkk
\nonumber \\&& \times
\Biggl(\ln 
\frac{k+mP/m_{\mbox{\scriptsize B}}+\sqrt{D}+i\nu/2a\sqrt{D}}
{k-mP/m_{\mbox{\scriptsize B}}-\sqrt{D}-i\nu/2a\sqrt{D}}
\nonumber \\
&&
%-\frac{2\pi m_{\mbox{\scriptsize B}}}{\hbar^2P} \int^{k_c}_{k_{\mbox{\scriptsize F}}} dkk 
+
\ln\frac{k+mP/m_{\mbox{\scriptsize B}}-\sqrt{D}-i\nu/2a\sqrt{D}}
{k-mP/m_{\mbox{\scriptsize B}}+\sqrt{D}+i\nu/2a\sqrt{D}}\Biggr).
\end{eqnarray} 
The integral can be solved \cite{PruBryMar98} 
to give

\begin{eqnarray}\lim_{k_c\to\infty}
{\cal I}&=&-\frac{8\pi m k_c}{\hbar^2}+\frac{4\pi m k_{\mbox{\scriptsize F}}}{\hbar^2}
\nonumber\\
&&+\frac{\pi}{\hbar^2}
\left(\frac{m_{\mbox{\scriptsize B}} k_{\mbox{\scriptsize F}}^2}{P}-\frac{m^2P}{m_{\mbox{\scriptsize B}}}-2m\sqrt{D}-\frac{m_{\mbox{\scriptsize B}} D}{P}\right)\nonumber\\
&&\times\ln
\frac{k_{\mbox{\scriptsize F}}+mP/m_{\mbox{\scriptsize B}}+\sqrt{D}+i\nu/2a\sqrt{D}}
{k_{\mbox{\scriptsize F}}-mP/m_{\mbox{\scriptsize B}}-\sqrt{D}-i\nu/2a\sqrt{D}}
\nonumber\\&&-\frac{\pi}{\hbar^2}
\left(\frac{m_{\mbox{\scriptsize B}} k_{\mbox{\scriptsize F}}^2}{P}-\frac{m^2P}{m_{\mbox{\scriptsize B}}}+2m\sqrt{D}-\frac{m_{\mbox{\scriptsize B}} D}{P}\right)\nonumber\\
&&\times\ln
\frac{k_{\mbox{\scriptsize F}}-mP/m_{\mbox{\scriptsize B}}+\sqrt{D}+i\nu/2a\sqrt{D}}
{k_{\mbox{\scriptsize F}}+mP/m_{\mbox{\scriptsize B}}-\sqrt{D}-i\nu/2a\sqrt{D}},
\label{first} 
\end{eqnarray}
where outside the logarithms we have taken the limit $\nu\to0$ (simply setting $\nu=0$),
and we have made use of the identity
\begin{equation}
\lim_{x\to \infty}
x^2\ln\frac{1+\alpha/x}{1-\alpha/x}= 2\alpha x,
\end{equation}
for the limit $k_c\to\infty$. There remains an ultraviolet divergent term;
the boson-fermion $T$-matrix [Eq.~(\ref{BSECM5})] is however ultimately renormalized by the second 
integral. 

The real part of $\cal{I}$ is readily evaluated in the limit $\nu\to0$ by setting $\nu=0$ and using the
absolute values inside the logarithms:
\begin{eqnarray}
\lim_{\nu\to 0}\mbox{Re}{\cal I}
&=&-\frac{8\pi m k_c}{\hbar^2}+\frac{4\pi m k_{\mbox{\scriptsize F}}}{\hbar^2}
\nonumber\\
&&+\frac{\pi}{\hbar^2}\left(\frac{m_{\mbox{\scriptsize B}} k_{\mbox{\scriptsize F}}^2}{P}
-\frac{m^2P}{m_{\mbox{\scriptsize B}}}-\frac{m_{\mbox{\scriptsize B}} D}{P}\right)
\nonumber\\
&&\times\ln\left|\frac{(k_{\mbox{\scriptsize F}}+mP/m_{\mbox{\scriptsize B}})^2-D}{(k_{\mbox{\scriptsize F}}-mP/m_{\mbox{\scriptsize B}})^2-D}\right|
\nonumber\\
&&-\frac{2\pi m}{\hbar^2}\sqrt{D}
\ln\left|\frac{(k_{\mbox{\scriptsize F}}+\sqrt{D})^2-(mP/m_{\mbox{\scriptsize B}})^2}
{(k_{\mbox{\scriptsize F}}-\sqrt{D})^2-(mP/m_{\mbox{\scriptsize B}})^2}\right|.
\nonumber \\
\end{eqnarray}

Using the identity (easily evaluated by polar decomposition)
\begin{equation}
\lim_{\nu\to0^+}\mbox{Im}\ln\frac{a+i\nu}{b-i\nu}=
\left\{\begin{array}{r@{\quad:\quad}l}
0 & \mbox{sgn}(a)=\mbox{sgn}(b)\\
\pi  & \mbox{sgn}(a)\neq\mbox{sgn}(b)\\
\end{array}\right.,
\end{equation} 
the imaginary part of  ${\cal I}$ in the limit $\nu\to 0$ can be evaluated to be:
\begin{equation}
\lim_{\nu\to 0}\mbox{Im}{\cal I} =
\frac{\pi^2}{\hbar^2}\left(\frac{m_{\mbox{\scriptsize B}} k_{\mbox{\scriptsize F}}^2}{P}
-\frac{m^2P}{m_{\mbox{\scriptsize B}}}-2m\sqrt{D}-\frac{m_{\mbox{\scriptsize B}} D}{P}\right),
\end{equation}
if $D>0$ and $k_{\mbox{\scriptsize F}}<|mP/m_{\mbox{\scriptsize B}}-\sqrt{D}|$;
\begin{equation}
\lim_{\nu\to 0}\mbox{Im}{\cal I} =-\frac{4\pi^2 m \sqrt{D}}{\hbar^2},
\end{equation}
if $D>0$ and 
$
|mP/m_{\mbox{\scriptsize B}}-\sqrt{D}|
<k_{\mbox{\scriptsize F}}<
mP/m_{\mbox{\scriptsize B}}+\sqrt{D}
$;
and 
\begin{equation}
\lim_{\nu\to 0}\mbox{Im}{\cal I} =0,
\end{equation}
 if $D\leq 0$ or 
$k_{\mbox{\scriptsize F}}>mP/m_{\mbox{\scriptsize B}}+\sqrt{D}$.

\subsection{The second integral ${\cal J}$}
We define 
\begin{eqnarray}
{\cal J}&=&-\int d^3{\mathbf{k}}\frac{1}
{\hbar^2{\mathbf{k}}_1^2/2m-\hbar^2{\mathbf{k}}^2/2m+i\nu}\nonumber\\
%&=&4\pi\int_0^{k_c}dk\frac{k^2}{ak^2-ak_1^2-i\nu}\nonumber\\
&=&\frac{4\pi}{a}\int_0^{k_c}dk\left(1-\frac{k_1^2+i\nu/a}{k^2-k_1^2-i\nu/a}\right),
\end{eqnarray}
where as before $a=\hbar^{2}/2m$, and  
 we have transformed to polar coordinates and integrated over the angle variables.
The integral can be evaluated \cite{PruBryMar98} to give
\begin{eqnarray}
{\cal J}&=&\frac{4\pi k_c}{a}
-\frac{2\pi}{a}\sqrt{k_1^2-i\nu/a}\ln\frac{k_c+\sqrt{k_1^2-i\nu/a}}{k_c-\sqrt{k_1^2-i\nu/a}}
\nonumber\\
&&+\frac{2\pi}{a}\sqrt{k_1^2-i\nu/a}\ln\frac{\sqrt{k_1^2-i\nu/a}}{-\sqrt{k_1^2-i\nu/a}}.
\end{eqnarray}
We then use
\begin{equation}
\lim_{k_{c}\to \infty}\ln\frac{k_c+\sqrt{k_1^2-i\nu/a}}{k_c-\sqrt{k_1^2-i\nu/a}}=0
\end{equation}
to get
\begin{eqnarray}
\lim_{\nu\to 0}{\cal J}&=&\frac{8\pi m k_c}{\hbar^2}
+i\frac{4\pi^2mk_1}{\hbar^2}.
\label{second}
\end{eqnarray}
If we now take the sum of Eqs.\ (\ref{first}) and (\ref{second}), the ultraviolet 
divergent terms cancel exactly. The resulting expression for ${\cal I} + {\cal J}$ can then be
substituted into Eq.\ (\ref{BSECM5}) to get Eq.\ (\ref{TMat}) for the
renormalized boson-fermion $T$-matrix.

\end{appendix}

%\end{fmffile}

\begin{references}
\bibitem{kf} 
This is only true for spin-polarized fermions. 
In general $k_{\mbox{\scriptsize F}}
=[6\pi^{2}n_{\mbox{\scriptsize F}}/(2s+1)]^{1/3}$,
where $s$ is the spin-degeneracy.
\bibitem{mod}
Considering Eq.\ (7.12) of Ref.\ \cite{FetWal71}, we 
take $V({\mathbf{x}},{\mathbf{x}}')_{\alpha\alpha',\beta\beta'}
=
U({\mathbf{x}},{\mathbf{x}}')\delta_{\alpha\alpha'}
\delta_{\beta\beta'}\delta_{\alpha(-\beta)}
+V({\mathbf{x}},{\mathbf{x}}')\delta_{\alpha\alpha'}
\delta_{\beta\beta'}\delta_{\alpha\beta}\delta_{\alpha(1/2)}
$, and
replace $\hat{\psi}_{1/2}({\mathbf{x}})$ by  $\hat{\phi}({\mathbf{x}})$ and
$\hat{\psi}_{-1/2}({\mathbf{x}})$ by 
$\hat{\Psi}({\mathbf{x}})$. The derivation is then the same, apart from some 
sign factors.
\bibitem{HPTcomment}
The proof
of the validity of the Hugenholtz-Pines theorem can be adopted litteraly from the pure boson 
case, since
it is based on how to replace condensate lines by non-condensate propagators; 
a procedure, which is unchanged in the present situation.
\end{references}
\end{document}